\documentclass[usenatbib]{mn2e}
\pdfoutput=1
\usepackage{amsmath}
\usepackage{amssymb}
\usepackage{graphicx}
\usepackage{rotating}
\usepackage{xspace}
\usepackage{url}
\usepackage{hyperref}
\usepackage{lscape}
\usepackage[varg]{txfonts}
\usepackage[T1]{fontenc}
\usepackage[usenames,dvipsnames]{color}

\def\tsys{$T_{\rm sys}$\xspace}
\def\D{\mathrm{d}}
\def\path{./figures}

\newcommand{\aap}{A\&A}

\newcommand{\mnras}{MNRAS}
\newcommand{\apjl}{ApJL}
\newcommand{\apjs}{ApJS}
\newcommand{\apj}{ApJ}
\newcommand{\aj}{ApJ}

\newcommand{\aapr}{A\&AR}
\newcommand{\pasp}{PASP}
\newcommand{\aaps}{A\&AS}

\title[Physical properties of LS~5039 through radio observations]{Physical properties of the gamma-ray binary LS~5039 through low and high frequency radio observations}
\author[]{
    B. Marcote$^{1,}$\thanks{E-mail:bmarcote@am.ub.es},
    M. Rib\'o$^{1,}$\thanks{Serra H\'unter Fellow.},
    J. M. Paredes$^{1}$ and C. H. Ishwara-Chandra$^{2}$\\
$^{1}$Departament d'Astronomia i Meteorologia, Institut de Ci\`encies del Cosmos, Universitat de Barcelona, IEEC-UB, Mart\'{\i} i\\ Franqu\`es 1, E08028 Barcelona, Spain\\
$^{2}$National Centre for Radio Astrophysics, TIFR, Post Bag 3, Ganeshkhind, 411007, Pune, India
}

\begin{document}

\date{Accepted 2015 April 24.  Received 2015 April 23; in original form 2015 February 23}

\pagerange{\pageref{firstpage}--\pageref{lastpage}} \pubyear{2015}

\maketitle

\label{firstpage}

\begin{abstract}
We have studied in detail the 0.15--15~GHz radio spectrum of the gamma-ray binary LS~5039 to look for a possible turnover and absorption mechanisms at low frequencies, and to constrain the physical properties of its emission. We have analysed two archival VLA monitorings, all the available archival GMRT data and a coordinated quasi-simultaneous observational campaign conducted in 2013 with GMRT and WSRT. The data show that the radio emission of LS~5039 is persistent on day, week and year timescales, with a variability $\la25~\%$ at all frequencies, and no signature of orbital modulation. The obtained spectra reveal a power-law shape with a curvature below 5~GHz and a turnover at $\sim0.5~\mathrm{GHz}$, which can be reproduced by a one-zone model with synchrotron self-absorption plus Razin effect. We obtain a coherent picture for a size of the emitting region of $\sim0.85~\mathrm{mas}$, setting a magnetic field of $B\sim20~\mathrm{mG}$, an electron density of $n_\mathrm{e}\sim4\times10^5~\mathrm{cm^{-3}}$ and a mass-loss rate of $\dot M\sim5\times10^{-8}~\mathrm{M_{\sun} yr^{-1}}$. These values imply a significant mixing of the stellar wind with the relativistic plasma outflow from the compact companion. At particular epochs the Razin effect is negligible, implying changes in the injection and the electron density or magnetic field. The Razin effect is reported for first time in a gamma-ray binary, giving further support to the young non-accreting pulsar scenario.
\end{abstract}

\begin{keywords}
binaries : close -- gamma rays: stars -- radio continuum: stars -- radiation mechanisms: non-thermal -- stars: individual: LS~5039 -- instrumentation: interferometers.
\end{keywords}

\section{Introduction}\label{sec:intro}

Gamma-ray binaries are extreme high-energy systems consisting of a young massive star and a compact object, which exhibit non-thermal emission from radio to very high energy gamma rays ($\ga 100~{\rm GeV}$). Their Spectral Energy Distribution (SED) is dominated by the MeV-GeV photons (see \citealt{dubus2013} for a recent review). 
Only a handful number of gamma-ray binaries have been discovered up to now: PSR~B1259$-$63 \citep{aharonian2005psr}, LS~5039 \citep{aharonian2005ls5039}, LS~I~+61~303 \citep{albert2006}, HESS~J0632+057 \citep{hinton2009,skilton2009}, and 1FGL~J1018.6$-$5856 \citep{fermi2012}. The first one is the only system which hosts a confirmed pulsar. The nature of the compact object still remains unknown for the rest of the gamma-ray binaries.

The behaviour of this kind of systems still represents a challenge for the high-energy astrophysical community. Although a microquasar scenario, typically seen in High Mass X-ray Binaries, was firstly proposed for these sources, the most recent results seem to favour a young non-accreting pulsar scenario for the four gamma-ray binaries deeply explored up to now \citep{dubus2013}.
Under the frame of this scenario, the broadband non-thermal emission of these extreme energetic systems is thought to arise from the particles which are accelerated in the shock between the wind of the compact object and the wind of the companion star. The gamma-ray emission would be produced by Inverse Compton upscattering of stellar photons by the most energetic electrons accelerated in the shock between the winds. The same population of electrons would produce the synchrotron emission observed at radio frequencies \citep{dubus2006,durant2011,bosch-ramon2009ls5039,bosch-ramon2013}.

LS~5039 (with IAU name J1826$-$1449 and coordinates $\alpha_{\rm J2000} = 18^{\rm h} 26^{\rm m} 15.06^{\rm s},\ \delta_{\rm J2000} = -14\degr 50\arcmin 54.3\arcsec$, \citealt{moldon2012}) is a gamma-ray binary system composed by a young O6.5\,V star and a compact object of 1--5 M$_{\sun}$, orbiting it every $\approx3.9~\mathrm{d}$ in an eccentric orbit with $e \approx 0.35$ \citep{casares2005}. The system is located at $2.9 \pm 0.8~{\rm kpc}$ from the Sun \citep{moldon2012}.
\citet{paredes2000} proposed for first time that LS~5039 was associated with an EGRET gamma-ray source. H.E.S.S. detected the binary system in the 0.1--4~TeV energy range \citep{aharonian2005ls5039}, and later reported variability at those energies along the orbital period \citep{aharonian2006}. {\em Fermi}/LAT confirmed the association with the EGRET source by clearly detecting the GeV counterpart, which showed also an orbital modulation \citep{abdo2009}. The system has been observed and detected with different X-ray satellites, showing moderate X-ray variability and a power-law spectrum with low X-ray photoelectric absorption consistent with interstellar reddening \citep{motch1997,ribo1999,reig2003,bosch-ramon2005,martocchia2005}. A clear orbital variability of the X-ray flux, correlated with the TeV emission, has been reported by \citet{takahashi2009}. Extended X-ray emission at scales up to 1--2\arcmin{} has also been detected \citep{durant2011}. X-ray pulsations have also been searched for, with null results \citep{rea2011}.

At radio frequencies, \citet*{marti1998} conducted a multifrequency study of LS~5039 with the Very Large Array (VLA) from 1.4 to 15~GHz with four observations spanning 3 months (one run per month). The radio emission of the source has a non-thermal origin, and can be described by a power-law with a spectral index of $\alpha = -0.46\pm 0.01$ (where the radio flux density at a given frequency, $S_\nu$, scales as $S_{\nu}\propto\nu^\alpha$). These authors reported a flux density variability below 30~\% with respect to the mean value.
\citet{ribo1999} and \citet{clark2001} presented the results of a radio monitoring campaign conducted on a daily basis during approximately one year with the Green Bank Interferometer (GBI) at 2.2 and 8.3~GHz frequencies. The emission was similar to the one reported in \citet{marti1998}, exhibiting a mean spectral index of $\alpha = -0.5_{-0.3}^{+0.2}$ \citep{ribo1999}. \citet{clark2001} constrained any orbital modulation to be $<4$~\% at 2.2~GHz. Searches of pulsed radio emission have been also conducted with null results, probably because of strong free-free absorption by the dense wind of the companion at the position of the compact object \citep{mcswain2011}.

The radio emission of the source is resolved at milliarcsecond scales using Very Long Baseline Interferometry (VLBI). \citet{paredes2000,paredes2002} found an asymmetrical bipolar extended emission on both sides of a bright dominant core, suggesting that LS~5039 was a microquasar. Later on, \citet{ribo2008} reported persistent emission from the dominant core and morphological changes of the extended emission on day timescales, which were difficult to reconcile with a microquasar model. Finally, \citet*{moldon2012ls5039} discovered morphological variability modulated with the orbital phase, and pointed out that a scenario with a young non-accreting pulsar is more plausible to explain the observed behaviour. In this case the relativistic wind from the pulsar shocks with the stellar wind of the massive companion and a cometary tail of accelerated particles forms behind the pulsar (see \citealt{moldon2012ls5039} and references therein for further details).

At low radio frequencies ($\la1~\mathrm{GHz}$) only few observations have been published up to now. \citet{pandey2007} performed two observations with the Giant Metrewave Radio Telescope (GMRT) simultaneously at 235 and 610~MHz, and reported a power-law spectrum with a spectral index of $\alpha\approx-0.8$ at these frequencies. On the other hand, \citet{godambe2008} conducted additional GMRT observations and reported a turnover at $\sim1~\mathrm{GHz}$, with a positive spectral index of $\alpha\approx+0.75$ at the same frequencies (further discussed in \citealt{bhattacharyya2012}). These results would indicate, surprisingly, that the flux variability of LS~5039 increases significantly below 1~GHz (see a possible explanation in the framework of the microquasar model in \citealt{bosch-ramon2009ls5039}). However, as discussed in this work, the change in the spectrum is the result of using different and uncomplete calibration procedures.

The radio spectrum of gamma-ray binaries is dominated by the synchrotron emission, which is expected to be self-absorbed at low frequencies. The spectrum could also reveal the presence of other absorption mechanisms such as free-free absorption or the Razin-Tsytovitch effect (also known as Tsytovitch-Eidman-Razin or simply Razin effect, \citealt{hornby1966}). This last mechanism, the suppression of synchrotron radiation in a plasma, is well known in solar bursts studies and widely reported in colliding wind binaries. These last systems probably share with gamma-ray binaries the origin of their emission: the collision between the winds of the two components of a binary system and the presence of shocks \citep{dougherty2003}.

Due to the surprising behaviour of the radio emission of LS~5039 at low frequencies obtained with a limited number of observations, we decided to make an in-depth study of the source. We have reduced and analysed two monitorings of LS~5039 at high frequencies that we conducted several years ago, most (if not all) of the publicly available low-frequency radio data, and conducted new coordinated observations to get a complete picture of the source behaviour at low and high radio frequencies. This has allowed us to monitor the variability of LS~5039 on different timescales (orbital and long-term), and determine its spectrum using simultaneous and non-simultaneous data.
This paper is organised as follows. In the next section we present the radio observations analysed in this work. In Sect.~\ref{sec:data-reduction} we detail the data reduction and analysis of these data, and the results are presented in Sect.~\ref{sec:results}. We detail the emission models used to describe the observed spectra in Sect.~\ref{sec:models}, and discuss the observed behaviour in Sect.~\ref{sec:discussion}. We summarise the obtained results and state the conclusions of this work in Sect.~\ref{sec:conclusions}.

\section{Radio observations}\label{sec:radio-observations}

\begin{figure}
        \includegraphics[width=0.51\textwidth]{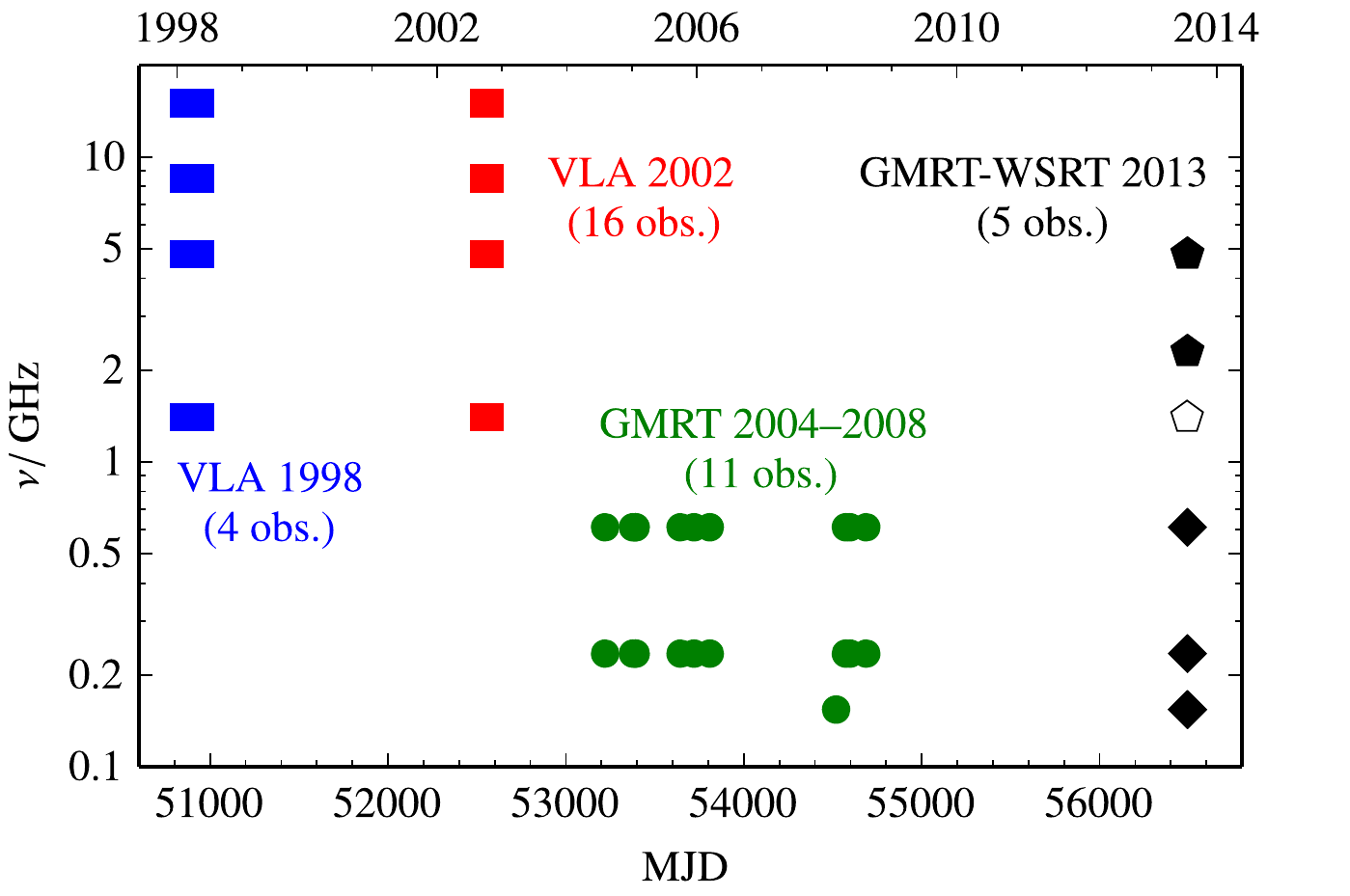}
        \caption{Summary of all the data presented in this work in a frequency versus time diagram, including the Modified Julian Date (MJD, bottom axis) and the calendar year (top axis). Squares represent the VLA monitorings in 1998 (blue) and 2002 (red), green circles correspond to the archival GMRT observations, black markers represent our coordinated GMRT (diamonds) and WSRT (pentagons) quasi-simultaneously campaign (the open pentagon represents a data set that could not be properly calibrated, see text). The radio data presented here span 15~yr.}
        \label{fig:summary-obs}
\end{figure}

The data presented along this work include two high-frequency VLA monitorings, all the available archival low-frequency radio observations performed with GMRT, and a new coordinated campaign that we conducted in 2013 using GMRT and the Westerbork Synthesis Radio Telescope (WSRT).\footnote{The VLA and ATCA data archives contain several isolated observations of LS~5039 above 1~GHz which have not been analysed in this work, because sporadic flux density measurements are not useful for the study of short term and orbital flux density modulations. These heterogeneous data sets could be useful in future spectral studies.}$^{,}$\footnote{We have also analysed two unpublished VLA observations conducted on 2006 December 16 and 18 at 330~MHz (project code AM877). However, we only obtained upper limits, because the noise is too high to clearly detect the source according to the GMRT data presented below. For this reason we will not explicitly include these observations and the obtained results in the rest of this work.}$^{,}$\footnote{We have analysed a LOFAR observation performed in 2011 at 150~MHz during its commissioning. However, the noise and the quality of the image at this stage were not accurate enough for our purposes.}
In Fig.~\ref{fig:summary-obs} we summarise these observations in a frequency versus time diagram. The details of all these observations, including the used facilities and configurations, the observation dates and frequencies, can be found in the first six columns of Table~\ref{tab:results}.

\subsection{VLA monitorings}   \label{sec:obs:vla}

Two multi-frequency monitorings have been conducted with the VLA: one in 1998 \citep{marti1998} and another one in 2002 (unpublished). In addition to the 2002 data set, we have reduced again the 1998 data set to guarantee that all the VLA data have been reduced using the same procedures.

The 1998 VLA monitoring consists on four observations conducted between February 11 and May 12 at 1.4, 4.8, 8.5 and 15~GHz using 2 Intermediate Frequency (IF) band pairs (RR and LL circular polarizations) of 50~MHz each (project code AP357). The amplitude calibrator used was 3C~286. The phase calibrators used were 1834$-$126 at 1.4~GHz, 1820$-$254 at 4.8 and 8.5~GHz, and 1911$-$201 at 15~GHz.

The 2002 VLA monitoring consists on 16 observations conducted between September 23 and October 21 (covering around 7 orbital cycles) at the same frequency bands and using the same calibrators than the previous monitoring (project code AP444).

\subsection{Archival GMRT observations}   \label{sec:obs:gmrt}

We have reduced all the existing archival GMRT data of LS~5039. These data set includes ten simultaneous observations at 235 and 610~MHz taken between 2004 and 2008: two published in \citet{pandey2007}, one reported in \citet{godambe2008}, and seven additional unpublished observations. There is an additional observation at 154~MHz performed in 2008. The GMRT data were obtained with a single IF, with dual circular polarization at 154~MHz (RR and LL), and single circular polarization at 235~MHz (LL) and 610~MHz (RR). The 154 and 235~MHz data were taken with a 8~MHz bandwidth and the 610~MHz data with 16~MHz (with 64 channels at 235~MHz and 128 channels at the other bands). In all these observations, 3C~286 and/or 3C~48 were used as amplitude calibrators and 1822$-$096 or 1830$-$360 as phase calibrators.  We have reduced again the published data to guarantee that all the GMRT data have been reduced using the same procedures.

\subsection{Coordinated GMRT-WSRT campaign in 2013}   \label{sec:obs:coord}

To study in detail the radio emission and absorption processes in LS~5039, we obtained for the first time a quasi-simultaneous spectrum of the source at low and high radio frequencies with GMRT (project code 24\_001) and WSRT (project code R13B012) on 2013 July 18--22.

We observed LS~5039 with GMRT at 154~MHz on July 18, 20, and 22 in three 8-hr runs, and simultaneously at 235 and 610~MHz on July 19 and 21 in two 4-hr runs. We conducted WSRT observations at 1.4, 2.3 and 4.8~GHz on July 19 and 21 in two 10-hr runs, which took place just after the GMRT ones at 235 and 610~MHz. The GMRT IFs, channels and polarizations are the same as the ones described in Sect.~\ref{sec:obs:gmrt}.
The total bandwidths are 16~MHz at 154 and 235~MHz, and 32~MHz at 610~MHz (double than in the earlier observations reported in Sect.~\ref{sec:obs:gmrt}). The WSRT data were obtained with 8 IFs of 46 channels each, with dual circular polarizations at 2.3~GHz (RR and LL) and dual linear polarizations at 1.4 and 4.8~GHz (XX and YY). The total bandwidth is 115~MHz at all frequencies.
We used 3C~286 and 3C~48 as amplitude calibrators for both observatories and 1822$-$096 as phase calibrator for GMRT (no phase calibrators are needed for WSRT due to the phase stability).

\begin{landscape}
\begin{table}
    \centering    
        \caption{Summary of all the data presented in this work. For each observation we show the facility used together with its array configuration (if any), the project code, the reference if previously published, the calendar date, the Modified Julian Date (MJD) of the central observation time, the corresponding orbital phase (using $P=3.90603~\mathrm{d}$ and $\mathrm{MJD_0} = 51942.59$), and the flux density values at each frequency with the 1-$\sigma$ uncertainty (3-$\sigma$ upper-limits in case of non-detection).}
        \label{tab:results}
\begin{tabular}{l@{~}c@{~}c@{~~~}c@{\hspace{10pt}}c@{\hspace{15pt}}cr@{}c@{}lr@{}c@{}lr@{}c@{}lr@{}c@{}lr@{}c@{}lr@{}c@{}lr@{}c@{}lr@{}c@{}l}
        \hline
        Facility & Project code & Ref. & Calendar Date & MJD & $\phi$ &
        \multicolumn{3}{c}{$S_{154\,\mathrm{MHz}}$} &
        \multicolumn{3}{c}{$S_{235\,\mathrm{MHz}}$} &
        \multicolumn{3}{c}{$S_{610\,\mathrm{MHz}}$} &
        \multicolumn{3}{c}{$S_{1.4\,\mathrm{GHz}}$} &
        \multicolumn{3}{c}{$S_{2.3\,\mathrm{GHz}}$} &
        \multicolumn{3}{c}{$S_{4.8\,\mathrm{GHz}}$} &
        \multicolumn{3}{c}{$S_{8.5\,\mathrm{GHz}}$} &
        \multicolumn{3}{c}{$S_{15\,\mathrm{GHz}}$}\\
        &&&&&&
        \multicolumn{3}{c}{$[\mathrm{mJy}]$} &
        \multicolumn{3}{c}{$[\mathrm{mJy}]$} &
        \multicolumn{3}{c}{$[\mathrm{mJy}]$} &
        \multicolumn{3}{c}{$[\mathrm{mJy}]$} &
        \multicolumn{3}{c}{$[\mathrm{mJy}]$} &
        \multicolumn{3}{c}{$[\mathrm{mJy}]$} &
        \multicolumn{3}{c}{$[\mathrm{mJy}]$} &
        \multicolumn{3}{c}{$[\mathrm{mJy}]$}\\
        \hline
        VLA-D & AP357 & (1) & 11/02/1998 & 50856.20 & 0.87 & &-& & &-& & &-& & 27 &$\pm$& 4 & &-& & 21.8 &$\pm$& 0.5 & 17.40 &$\pm$& 0.10 & 12.4 &$\pm$& 0.3\\
VLA-A & AP357 & (1) & 10/03/1998 & 50883.10 & 0.76 & &-& & &-& & &-& & 38.6 &$\pm$& 1.0 & &-& & 23.70 &$\pm$& 0.10 & 17.77 &$\pm$& 0.08 & 13.3 &$\pm$& 0.2\\
VLA-A & AP357 & (1) & 09/04/1998 & 50913.00 & 0.41 & &-& & &-& & &-& & 39.8 &$\pm$& 0.7 & &-& & 22.70 &$\pm$& 0.10 & 15.85 &$\pm$& 0.07 & 11.80 &$\pm$& 0.18\\
VLA-A & AP357 & (1) & 12/05/1998 & 50946.00 & 0.86 & &-& & &-& & &-& & 45.5 &$\pm$& 1.0 & &-& & 26.0 &$\pm$& 0.2 & 18.23 &$\pm$& 0.12 & 13.9 &$\pm$& 0.2\\
VLA-BC& AP444 & - & 23/09/2002 & 52540.96 & 0.19 & &-& & &-& & &-& & 36.8 &$\pm$& 1.4 & &-& & 20.5 &$\pm$& 0.5 & 14.3 &$\pm$& 0.5 & 9.4 &$\pm$& 0.7\\
VLA-BC& AP444 & - & 28/09/2002 & 52545.07 & 0.24 & &-& & &-& & &-& & &-& & &-& & 24.4 &$\pm$& 0.2 & 16.5 &$\pm$& 0.3 & &-&\\
VLA-BC& AP444 & - & 29/09/2002 & 52546.96 & 0.73 & &-& & &-& & &-& & 29.2 &$\pm$& 1.7 & &-& & 22.3 &$\pm$& 0.4 & 14.7 &$\pm$& 0.3 & 8.4 &$\pm$& 0.7\\
VLA-BC& AP444 & - & 02/10/2002 & 52549.04 & 0.26 & &-& & &-& & &-& & 27.5 &$\pm$& 1.3 & &-& & 15.6 &$\pm$& 0.5 & 9.7 &$\pm$& 0.5 & 6.5 &$\pm$& 0.6\\
VLA-BC& AP444 & - & 05/10/2002 & 52552.03 & 0.03 & &-& & &-& & &-& & 33.6 &$\pm$& 1.2 & &-& & 18.8 &$\pm$& 0.3 & 11.3 &$\pm$& 0.4 & 8.4 &$\pm$& 0.4\\
VLA-BC& AP444 & - & 07/10/2002 & 52554.98 & 0.78 & &-& & &-& & &-& & 32 &$\pm$& 2 & &-& & 15.6 &$\pm$& 0.6 & 6.3 &$\pm$& 0.5 & &-&\\
VLA-C & AP444 & - & 12/10/2002 & 52559.01 & 0.81 & &-& & &-& & &-& & 34.5 &$\pm$& 1.9 & &-& & 20.0 &$\pm$& 0.4 & 16.7 &$\pm$& 0.3 & 13.3 &$\pm$& 0.4\\
VLA-C & AP444 & - & 12/10/2002 & 52559.93 & 0.05 & &-& & &-& & &-& & 38.5 &$\pm$& 1.1 & &-& & 25.6 &$\pm$& 0.2 & 18.0 &$\pm$& 0.2 & 12.0 &$\pm$& 0.5\\
VLA-C & AP444 & - & 14/10/2002 & 52561.03 & 0.33 & &-& & &-& & &-& & &-& & &-& & 20.25 &$\pm$& 0.19 & 14.5 &$\pm$& 0.3 & 10.8 &$\pm$& 0.5\\
VLA-C & AP444 & - & 15/10/2002 & 52562.02 & 0.58 & &-& & &-& & &-& & &-& & &-& & 25.1 &$\pm$& 0.2 & 16.5 &$\pm$& 1.0 & 10 &$\pm$& 2\\
VLA-C & AP444 & - & 15/10/2002 & 52562.98 & 0.83 & &-& & &-& & &-& & 42 &$\pm$& 2 & &-& & 29.6 &$\pm$& 0.2 & 20.8 &$\pm$& 0.3 & 15.3 &$\pm$& 0.6\\
VLA-C & AP444 & - & 17/10/2002 & 52564.96 & 0.34 & &-& & &-& & &-& & 31.6 &$\pm$& 1.7 & &-& & 21.6 &$\pm$& 0.2 & 14.9 &$\pm$& 0.3 & 9.3 &$\pm$& 0.7\\
VLA-C & AP444 & - & 19/10/2002 & 52566.01 & 0.60 & &-& & &-& & &-& & 35.0 &$\pm$& 1.8 & &-& & 23.6 &$\pm$& 0.5 & 13.0 &$\pm$& 0.6 & 7.9 &$\pm$& 0.7\\
VLA-C & AP444 & - & 19/10/2002 & 52566.91 & 0.83 & &-& & &-& & &-& & 36.4 &$\pm$& 1.3 & &-& & 23.9 &$\pm$& 0.2 & 16.6 &$\pm$& 0.2 & 3.7 &$\pm$& 0.3\\
VLA-C & AP444 & - & 20/10/2002 & 52567.95 & 0.10 & &-& & &-& & &-& & 26.2 &$\pm$& 1.5 & &-& & 18.9 &$\pm$& 0.2 & 14.8 &$\pm$& 0.2 & 11.2 &$\pm$& 0.6\\
VLA-C & AP444 & - & 21/10/2002 & 52568.94 & 0.35 & &-& & &-& & &-& & 29.1 &$\pm$& 1.6 & &-& & 24.1 &$\pm$& 0.3 & 18.6 &$\pm$& 0.3 & 12.3 &$\pm$& 0.4\\
GMRT & 06PDA01 & (2) & 03/08/2004 & 53220.57 & 0.18 & &-& & \multicolumn{3}{l}{$<200$} & 51 &$\pm$& 5 & &-& & &-& & &-& & &-& & &-&\\
GMRT & 07PDA01 & (2) & 07/01/2005 & 53377.21 & 0.28 & &-& & 27 &$\pm$& 9 & 28 &$\pm$& 4 & &-& & &-& & &-& & &-& & &-&\\
GMRT & 07PDA01 & - & 21/01/2005 & 53391.20 & 0.86 & &-& & \multicolumn{3}{l}{$<11$}& 38 &$\pm$& 3 & &-& & &-& & &-& & &-& & &-&\\
GMRT & 07PDA01 & - & 22/01/2005 & 53392.20 & 0.12 & &-& & 18 &$\pm$& 6 & 49 &$\pm$& 3 & &-& & &-& & &-& & &-& & &-&\\
GMRT & 08SKC01 & - & 01/10/2005 & 53644.60 & 0.75 & &-& & 28 &$\pm$& 9 & 46 &$\pm$& 3 & &-& & &-& & &-& & &-& & &-&\\
GMRT & 09PDA01 & - & 16/12/2005 & 53720.17 & 0.09 & &-& & 26 &$\pm$& 8 & 41 &$\pm$& 2 & &-& & &-& & &-& & &-& & &-&\\
GMRT & 09SBB01 & (3, 4) & 15/03/2006 & 53809.04 & 0.84 & &-& & 31 &$\pm$& 9 & 49 &$\pm$& 3 & &-& & &-& & &-& & &-& & &-&\\
GMRT & 13MPA01 & - & 23/02/2008 & 54519.00 & 0.63 & &$<6$& & &-& & &-& &&-& & &-& & &-& & &-& & &-&\\
GMRT & 14SVG02 & - & 17/04/2008 & 54573.81 & 0.66 & &-& & 19 &$\pm$& 6 & 35 &$\pm$& 2 & &-& & &-& & &-& & &-& & &-&\\
GMRT & 14SVG02 & - & 10/05/2008 & 54596.78 & 0.55 & &-& & 25 &$\pm$& 8 & 31.3 &$\pm$& 1.8 & &-& & &-& & &-& & &-& & &-&\\
GMRT & 14SVG02 & - & 10/08/2008 & 54688.63 & 0.02 & &-& & \multicolumn{3}{l}{$<7$} & 40 &$\pm$& 2 & &-& & &-& & &-& & &-& & &-&\\
GMRT & 24\_001 & - & 18/07/2013 & 56491.56 & 0.60 & &$<6$& & &-& & &-& & &-& & &-& & &-& & &-& & &-&\\
GMRT & 24\_001 & - & 19/07/2013 & 56492.60 & 0.87 & &-& & 34 &$\pm$& 11 & 51 &$\pm$& 3 & &-& & &-& & &-& & &-& & &-&\\
WSRT & R13B012 & - & 19/07/2013 & 56492.75 & 0.90 & &-& & &-& && -&& &- & & 30.6 &$\pm$& 1.0 & 22.42 &$\pm$& 0.15 & &-& & &-&\\
GMRT & 24\_001 & - & 20/07/2013 & 56493.56 & 0.11 & &$<12$& & &-& & &-& & &-& & &-& & &-& & &-& & &-&\\
GMRT & 24\_001 & - & 21/07/2013 & 56494.57 & 0.37 & &-& & 24 &$\pm$& 7 & 37 &$\pm$& 2 & &-& & &-& & &-& & &-& & &-&\\
WSRT & R13B012 & - & 21/07/2013 & 56494.75 & 0.42 & &-& & &-& & &-& & &-& & 27.2 &$\pm$& 0.7 & 18.68 &$\pm$& 0.11 & &-& & &-&\\
GMRT & 24\_001 & - & 22/07/2013 & 56495.73 & 0.67 & &$<8$& & &-& & &-& & &-& & &-& & &-& & &-& & &-&\\

        \hline
        \end{tabular}
        
        \medskip{(1)~\citet{marti1998}; (2)~\citet{pandey2007}; (3)~\citet{godambe2008}; (4)~\citet{bhattacharyya2012}}
\end{table}
\end{landscape}
\section{Data reduction and analysis}\label{sec:data-reduction}

All the radio data have been calibrated and analysed using standard procedures
mainly within AIPS\footnote{The NRAO Astronomical Image Processing System.\\
\url{http://www.aips.nrao.edu}}.
Obit\footnote{\url{http://www.cv.nrao.edu/~bcotton/Obit.html}} \citep{cotton2008},
Parseltongue\footnote{\url{http://www.jive.nl/jivewiki/doku.php?id=parseltongue:parseltongue}} \citep{kettenis2006}
and SPAM\footnote{\url{https://safe.nrao.edu/wiki/bin/view/Main/HuibIntemaSpam}} \citep{intema2009} have also been used mainly to run scripts that call AIPS tasks to reduce the data.
The raw visibilities were loaded into AIPS and flagged removing telescope off-source times, instrumental problems or radio frequency interferences (RFI). The amount of data flagged due to RFI depends strongly on the frequency. Whereas at frequencies above 1.4~GHz we do not observe a relevant amount of RFI, at 150~MHz we lose an important fraction of the data in some observations, even up to 30~\%.

The VLA data have been reduced using amplitude and phase calibration steps. Self-calibration on the target source was successful at 1.4, 4.8 and 8.5~GHz, but failed for most of the 15~GHz observations due to the faintness of the source. For this reason, we preferred not to self-calibrate any VLA data of LS~5039 to compare the results at all frequencies with exactly the same data reduction process.

The GMRT data have been reduced as follows. First, we performed an a priori amplitude and phase calibration using an RFI-free central channel of the band. Afterwards, we performed the bandpass calibration. Finally, we calibrated all the channels of the band in amplitude and phase considering the bandpass calibration.
We imaged the target source and conducted several cycles of self-calibration/imaging to correct for phase and amplitude errors.
Given the huge field of view of the GMRT images (few degrees), we considered not only the $uv$-plane but the full $uvw$-space during the imaging process \citep{thompson1999}.

In addition, for the GMRT data we have also performed a correction of the system temperature, $T_{\mathrm{sys}}$, for each antenna to subtract the contribution of the Galactic diffuse emission, which is relevant at low frequencies.
This $T_{\mathrm{sys}}$ correction was performed using archival self-correlated observations of the target source and the calibrators, but also using recent data taken at our request after the campaign in 2013. A detailed explanation of the system temperature correction process for the GMRT data can be found in Appendix~\ref{app:tsys}. The obtained \tsys corrections are directly applied to the flux densities of the final target images. We note that these corrections imply an additional source of uncertainty.

The WSRT data have been reduced using amplitude and phase calibration steps. The solutions for the amplitude calibrators were directly extrapolated to the target source. The data were imaged and self-calibrated a few times. We had calibration problems at 1.4~GHz, and we could not extract any result from the corresponding data.

During the imaging process we have used a Briggs robustness parameter of zero in all cases. For the VLA and GMRT data the synthesized beam had a major axis around 2 times the minor axis. For the VLA in A configuration we have obtained geometrical mean values for the synthesized beam between $2.3\arcsec$ (1.4~GHz) and $0.18\arcsec$ (15~GHz). In D configuration these values increase up to $58\arcsec$ at 1.4~GHz. With GMRT we have obtained synthesized beams between $18\arcsec$ and $5\arcsec$ (at 154~MHz and 610~MHz, respectively). The WSRT observations exhibit an elongated synthesized beam with a major axis around 10 times the minor axis as a consequence of the linear configuration of the array and the declination of the source. For WSRT the synthesized beam was $110\arcsec\times10\arcsec$ in PA = $5\degr$ at 2.3~GHz, and $48\arcsec\times5.7\arcsec$ in PA = $-1\degr$ at 4.8~GHz.

The measurement of the flux density value of LS~5039 in each image has been conducted using the {\tt tvstat} task of AIPS, by considering a small region centered around the target.  We also measured the root-mean-square (rms) flux density of a region centered on the source, but clearly excluding it, and considered this to be the flux density uncertainty. However, in some cases the noise around the source is larger than the obtained rms (due to poor $uv$-plane coverage, calibration errors, etc.). In such cases, we repeated the flux density measurement a few times using different region sizes, and considered the dispersion of the obtained values as a measure of the flux density uncertainty. In any case, these uncertainties were never above 50~\% of the corresponding rms values. The {\tt jmfit} task of AIPS has also been used to double check the flux densities of LS~5039.

To guarantee the reliability of the measured flux densities, we have monitored the flux densities of the amplitude calibrator for the VLA observations and the flux densities of several background sources detected in the field of view of LS~5039 for the GMRT observations. The constancy of the obtained VLA flux densities (below 3~\% at 1.4 and 15~GHz and below 0.4~\% at 4.8 and 8.5~GHz) and the lack of trends in the GMRT flux densities, allow us to ensure that the reported variability of LS~5039 (see below) is caused by the source itself, and not by additional effects during the observations or the calibration process.

\section{Results}\label{sec:results}

LS~5039 appears as a point-like source in all the VLA, GMRT and WSRT images, and we summarise the results for all these observations in Table~\ref{tab:results}. In this section we focus first on the VLA monitoring performed in 2002 to study the variability of LS~5039 at high frequencies along consecutive orbital cycles. Later on, we move to the GMRT observations to study the variability at low frequencies. We continue by reporting the high frequency spectra from the VLA monitoring of 2002. We then combine all the data to show the full radio spectrum of LS~5039 from 0.15 to 15~GHz. Finally, we report on the quasi-simultaneous spectra from 0.15 to 5~GHz from the coordinated GMRT-WSRT campaign in 2013.

\subsection{Light-curve at high frequencies $\bmath{(>1~\mathrm{GHz})}$}

\begin{figure}
        \includegraphics[width=0.48\textwidth]{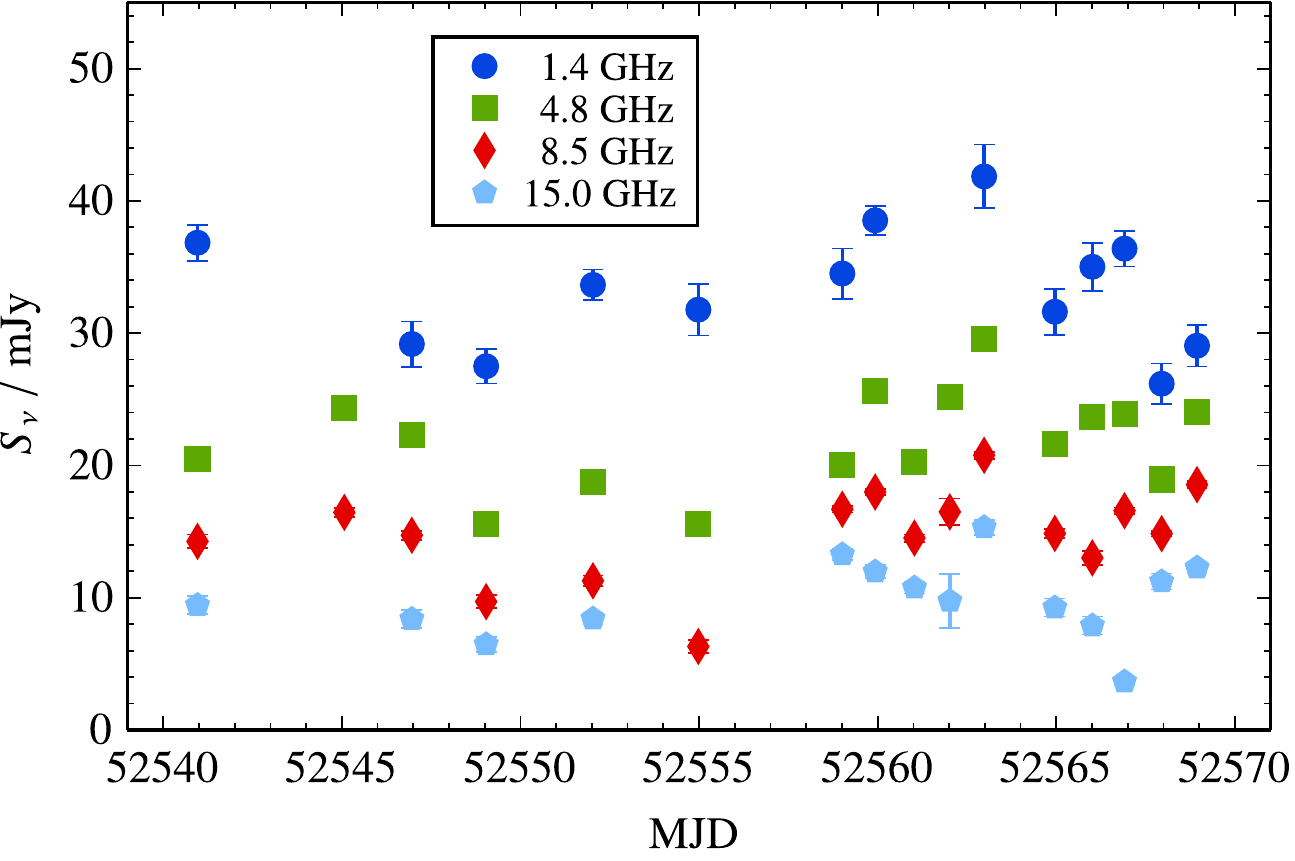}
        \caption{Multi-frequency light-curves of LS~5039 as a function of the Modified Julian Date (MJD) for the VLA monitoring in 2002. Error bars represent 1-$\sigma$ uncertainties.}
        \label{fig:vla-mjd-flux}
\end{figure}
Figure~\ref{fig:vla-mjd-flux} shows the multi-frequency light-curves obtained from the VLA monitoring in 2002 as a function of time. The flux of LS~5039 is persistent, and shows variability along all the observing period ($\approx 28~\mathrm{d}$) on timescales as short as one day. The trend of the variability is roughly similar at all frequencies. Apart from the daily variability, there is an evolution of the flux density on timescales of the order of 10 days. A $\chi^2$ test provides a probability of variability of 8--40$\sigma$, depending on the frequency. Since the flux density values encompass those of the 1998 monitoring (see \citealt{marti1998} and our values in Table~\ref{tab:results}) and show similar average values at all frequencies, we conclude that the emission of LS~5039 is similar in both epochs (1998 and 2002).

\begin{figure}
        \includegraphics[width=0.483\textwidth]{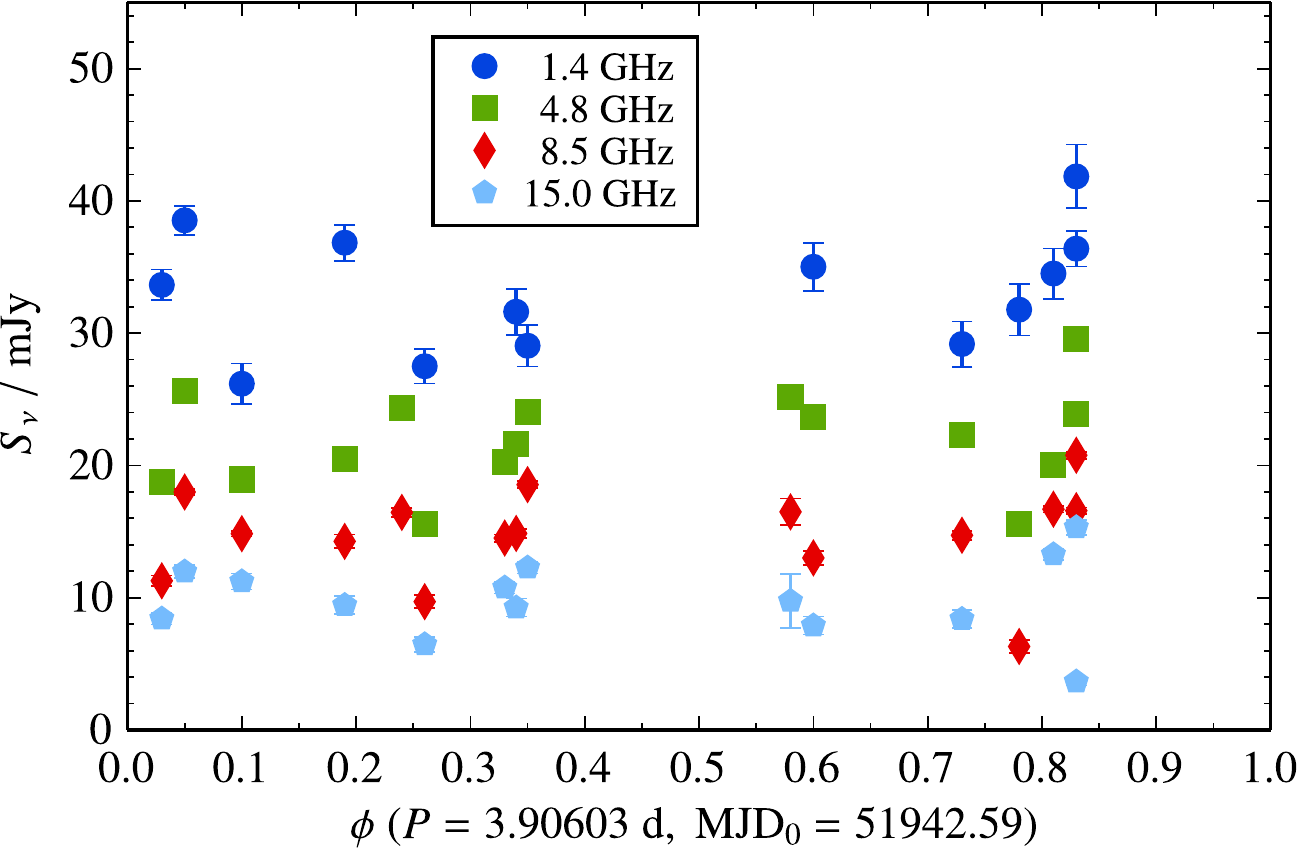}
        \caption{Multi-frequency light-curves of LS~5039 as a function of the orbital phase for the VLA monitoring in 2002. The campaign covers around 7 consecutive orbital cycles. Error bars represent 1-$\sigma$ uncertainties.}
        \label{fig:vla-phase-flux}
\end{figure}

In Fig.~\ref{fig:vla-phase-flux} we plot the same data as a function of the orbital phase. The reduced number of observations and the presence of some gaps ($\sim$0.35--0.55, $\sim 0.60$--$0.70$, $\sim 0.85$--$1.00$) does not allow us to clearly report on any significant orbital modulation on top of the reported daily and weekly variability. We note that at orbital phase $\sim 0.8$ there is a large dispersion of flux densities at all frequencies.
We have also binned the multi-frequency light-curves with different bin sizes (0.15, 0.20, 0.25) and different central bin positions to increase the statistics, but we have not seen any significant deviation with respect to the mean values or larger variabilities at specific orbital phases (not even at phase 0.8).
To further search for deviations at phase $\sim 0.8$, we have re-analysed the full GBI data at 2.2 and 8.3~GHz, consisting on 284 observations spanning 340 days \citep{ribo1999,clark2001}, obtaining null results.

\begin{figure}
        \includegraphics[width=0.48\textwidth]{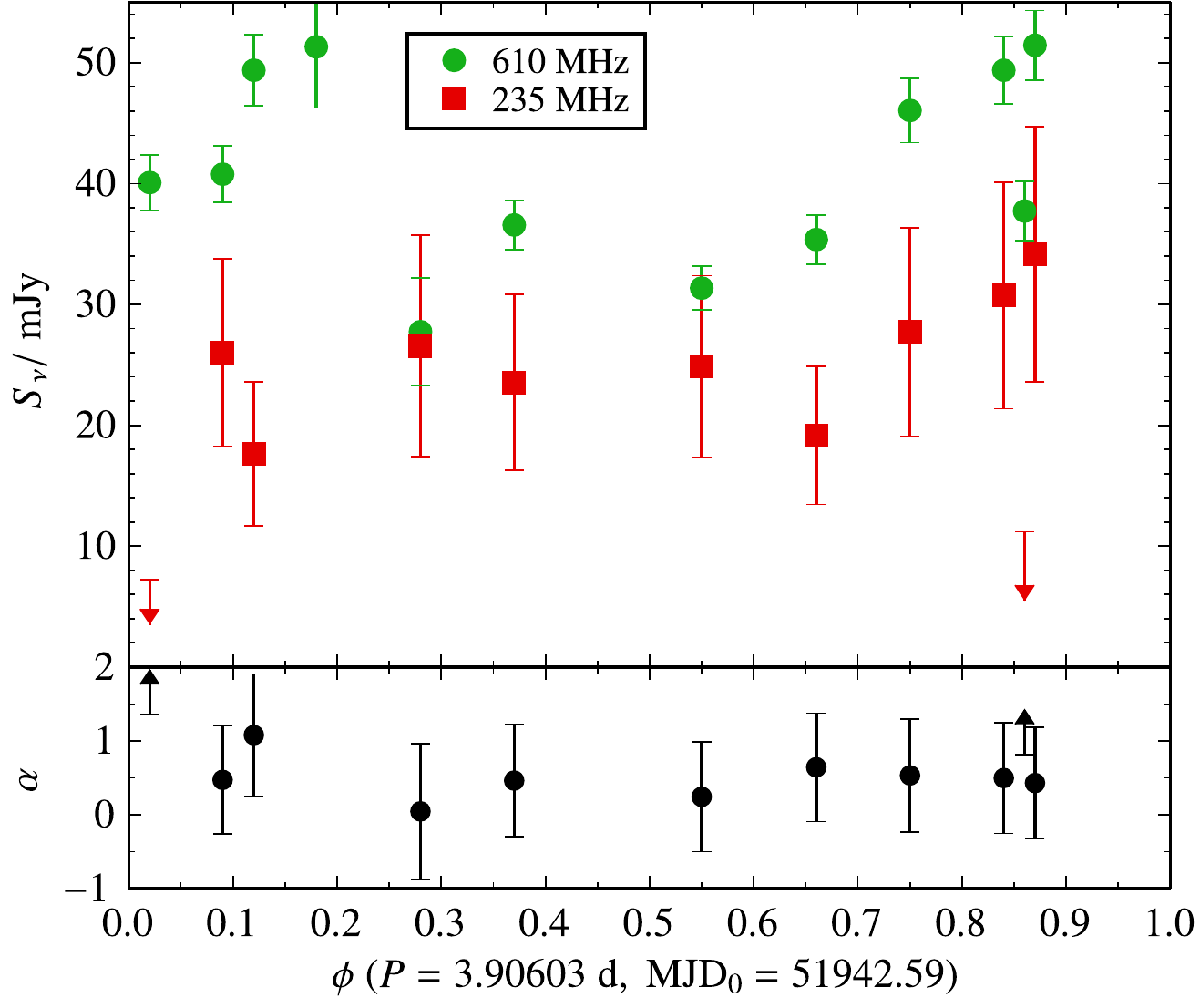}
        \caption{{\em Top:} light-curves of LS~5039 as a function of the orbital phase at 235 and 610~MHz of the archival GMRT observations from 2004 to 2008 and the two GMRT observations from the coordinated GMRT-WSRT campaign in 2013. Error bars represent 1-$\sigma$ uncertainties (resulting from the combination of the image noise and the \tsys correction uncertainties). Upper limits at 235~MHz are plotted at the 3-$\sigma$ level. {\em Bottom:} spectral indexes derived from the data above. We observe a mean spectral index of $\alpha \approx 0.5$ between 235 and 610~MHz.}
        \label{fig:gmrt-phase-flux}
\end{figure}

\subsection{Light-curve at low frequencies $\bmath{(<1~\mathrm{GHz})}$}

Figure~\ref{fig:gmrt-phase-flux}-top shows the light curves of LS~5039 at 235 and 610~MHz as a function of the orbital phase from all the analysed GMRT observations: the archival GMRT observations (2004--2008) and the two GMRT observations from the coordinated GMRT-WSRT campaign in 2013. The source exhibits a persistent radio emission at both frequencies during this 9-year period. At 610~MHz LS~5039 shows clear variability ($6\sigma$). At 235~MHz the variability is not significant ($1\sigma$) for the most conservative decision of considering the upper-limits as detections with flux densities corresponding to 3 times the uncertainties.

These smaller significances in the variability with respect to the high frequency ones are produced mainly by the larger uncertainties of the flux densities, which are a combination of the image noise and the dominant \tsys correction uncertainties.
When we apply these corrections, we increase the typical 1--3~\% uncertainties to $\approx 8$~\% uncertainties in the flux densities at 610~MHz, and from 6--16~\% to $\approx 40$~\% at 235~MHz.
As the \tsys corrections are scaling factors which are equally applied to all the observations and only depend on the frequency, they only affect the absolute values of the flux densities, but not their relative differences. This means that we can determine the dispersion of these data prior to applying the \tsys corrections. In this case, we obtain that the source is variable at 610~MHz and 235~MHz with a confidence level of $35\sigma$ and $6\sigma$, respectively. At 610~MHz variability on day timescales is also detected (see MJD 53391--53392 and MJD 56492--56494 in Table~\ref{tab:results}).

Therefore, we see the same behaviour than at high frequencies: variability on timescales as short as one day, with a persistent emission along the years, and without clear signals of being orbitally modulated. Observing the 610~MHz data (Fig.~\ref{fig:gmrt-phase-flux}-top), a small orbital modulation seems to emerge, with a broad maximum located at $\phi \sim 0.8$--$0.2$ and a broad minimum at $\phi \sim 0.3$--$0.7$.
To guarantee that this variability is not due to calibration issues, we have monitored the flux densities of up to 7 compact background sources in the field of LS~5039. Their flux densities vary in different ways, but none of them shows the same trend of variability as LS~5039.
Whether the orbital modulation suggested above for LS~5039 is produced by the poor sampling or by an intrinsic modulation of the source will remain unknown until further detailed observations are conducted.

The spectral indexes from Fig.~\ref{fig:gmrt-phase-flux}-bottom provide a mean of $\alpha = 0.5 \pm 0.8$. Given the large uncertainties, we can not claim for variability at more than a 3-$\sigma$ level.
However, if we do not consider the uncertainties related to the \tsys corrections we do obtain a significant variability of the spectral index, which is exclusively produced by the two observed upper-limits.

\begin{figure*}
	\centering
        \includegraphics[width=10cm]{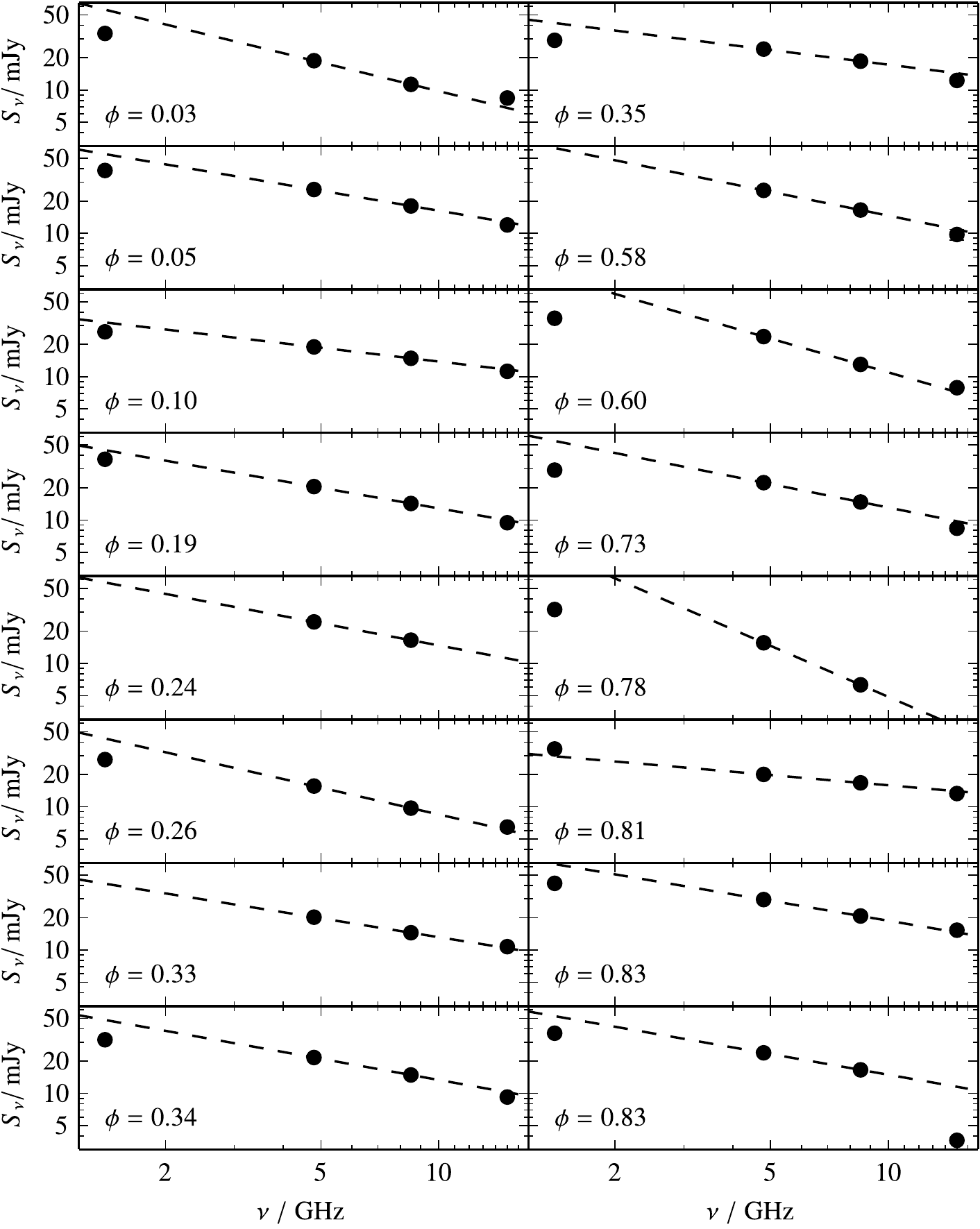}
        \caption{Spectra of LS~5039 from the VLA monitoring in 2002 ordered with the orbital phase. The dashed line represents the power law derived from the 4.8 and 8.5~GHz data. Signals of a curved spectrum below $\sim2~{\rm GHz}$ are visible in most cases. At orbital phases $\sim0.8$ we observe the most extreme spectra (the steepest and the flattest ones).}
        \label{fig:vla-freq-flux}
\end{figure*}

\subsection{High frequency spectra $\bmath{(>1~\mathrm{GHz})}$}

We show in Fig.~\ref{fig:vla-freq-flux} the spectrum of LS~5039 for each observation within the VLA monitoring in 2002. The spectra can be roughly fitted with a power law, but they show a slight curvature below $\sim5~{\rm GHz}$ in most cases.
Given this curvature and the fact that at 15~GHz we can have a larger dispersion due to changing weather conditions, we have derived a power law for the remaining frequencies (4.8 and 8.5~GHz) to estimate the spectral index $\alpha$. The obtained values are in the range from $-0.35$ to $-1.1$, with an average value of $\alpha = -0.57\pm 0.12$, compatible with previously reported values. No signatures of orbital modulation are detected. However, around orbital phase 0.8 we observe the most extreme spectra, with the steepest and flattest ones.

\subsection{Non-simultaneous spectrum}

\begin{figure}
        \includegraphics[width=0.475\textwidth]{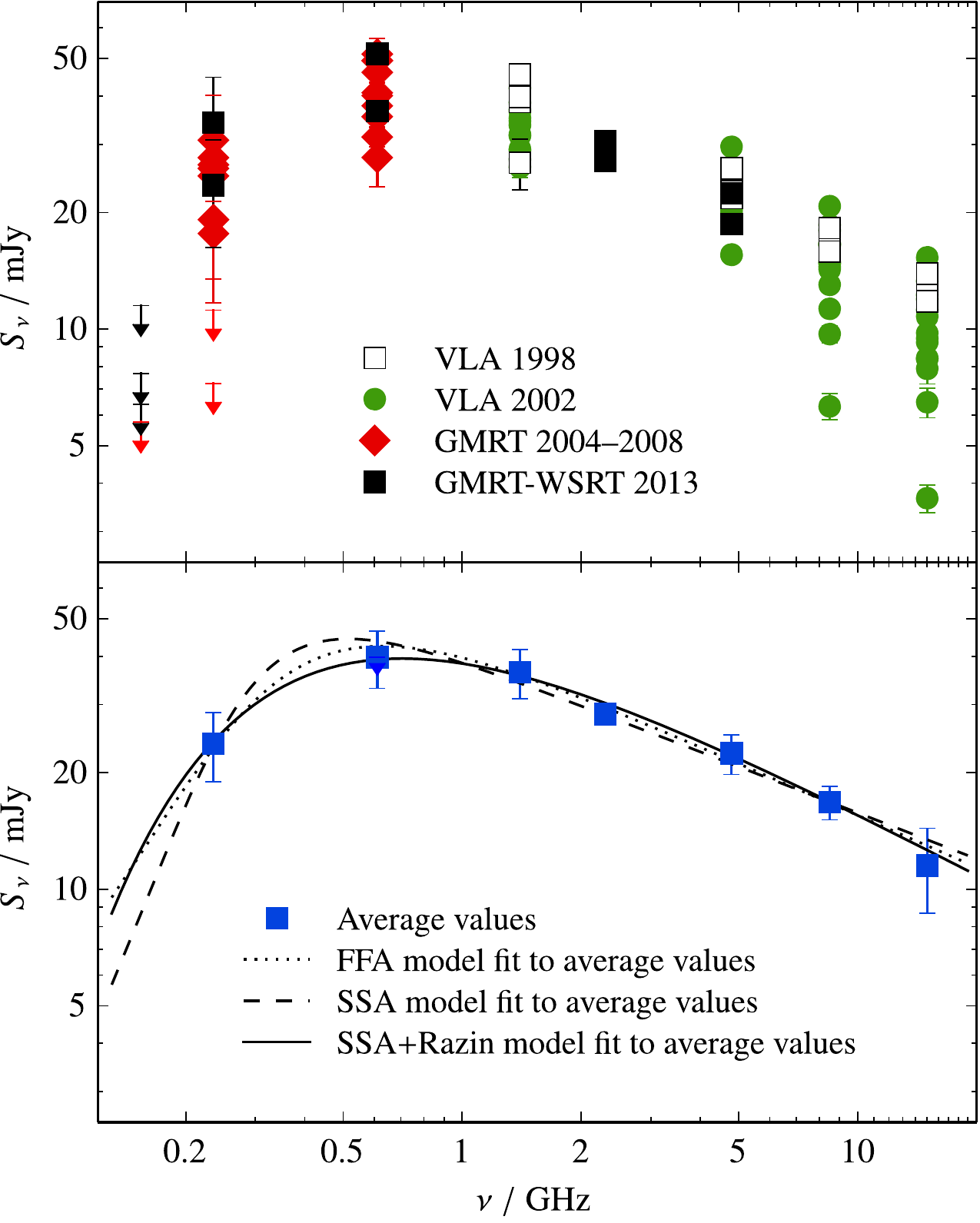}
        \caption{{\em Top}: Non-simultaneous spectrum of LS~5039 obtained with all the data presented in this work (VLA 1998 \& 2002, GMRT 2004--2008, GMRT-WSRT 2013). Error bars represent 1-$\sigma$ uncertainties. Upper limits are plotted at the 3-$\sigma$ level (one at 235~MHz lies outside the figure limits).
        {\em Bottom}: Average flux density value at each frequency. Error bars represent the standard deviations. The upper-limit at 235~MHz outside the figure limits has not been considered in the averaging. This average spectrum has been fitted with three different absorption models: free-free absorption (FFA), synchrotron self-absorption (SSA), and SSA plus Razin effect (SSA+Razin). Mean square errors (within the symbol sizes, not shown in the figure) have been used as uncertainties in the fitting process (see text), while the data at 2.3~GHz and the upper-limits at 154~MHz have not been considered.}
        \label{fig:all-freq-flux}
\end{figure}

Figure~\ref{fig:all-freq-flux}-top shows the obtained spectrum (0.15--15~GHz) from LS~5039 with all the data presented in this work (VLA monitorings in 1998 and 2002, the archival GMRT observations in 2004--2008, and the coordinated GMRT-WSRT campaign in 2013). All these data cover around 15 years of observations. The source is always detected above 0.5~GHz. At 235~MHz the source is detected most of the times, but as already mentioned there are some upper limits. On 2004 August 3 ($\phi=0.18$) the rms of the image at this frequency is much higher than the flux density expected for the source and thus these data have not been taken into account in what follows. At 154~MHz the source is not detected in any observation, neither combining all the existing data. The spectrum is clearly curved, with a turnover around $\sim 0.5~\mathrm{GHz}$.

We observe that the emission of LS~5039 along the years remains persistent with a moderate variability above the turnover frequency (below the turnover the uncertainties become larger). Given this behaviour, we averaged the flux density of LS~5039 at each frequency, considering the 235~MHz upper-limits as detections with flux densities corresponding to 3 times the uncertainties. We show the average values in Fig.~\ref{fig:all-freq-flux}-bottom and quote them in Table~\ref{tab:mean-values}.
We observe a standard deviation below 25~\% at all frequencies. The average value and the variability reported at 2.3~GHz (from WSRT data) are not as accurate as in the other cases because we only have two observations at such frequency.

\begin{table}
	\centering
        \caption{Average value of the flux density ($S_{\nu}$) of LS~5039 and its relative variability ($\delta S_{\nu}\cdot S_{\nu}^{-1}$) at each frequency, considering all the data presented in this work. $N_{\rm obs}$ values have been considered at each frequency. In all cases we report a variability below 25~\%. At 2.3~GHz we only have two observations to compute the mean value, so the amplitude of the variability is not relevant and can not be compared with the rest of the data.}
        \begin{tabular}{c@{\hspace{+25pt}}c@{\hspace{+25pt}}r@{}c@{}l@{}@{\hspace{+25pt}}c}
            \hline
            $\nu\,/\ \mathrm{GHz}$ & $N_{\rm obs}$ & \multicolumn{3}{l}{$S_{\nu} /\ \mathrm{mJy}$} &
            $\delta S_{\nu}\cdot S_{\nu}^{-1} /\ \%$\\
            \hline
            0.235& 11 &23& $\,\pm\,$ &5 & 20\\
0.610& 12 &39& $\pm$ &6 & 16\\
1.4  & 17 & 36& $\pm$ &5 & 15\\
2.3  & 2  & 28.3& $\pm$ &1.6 & 6\\
4.8  & 22 & 22& $\pm$ &3 & 12\\
8.5  & 20 & 16.8& $\pm$ &1.7 & 10\\
15   & 18 & 11& $\pm$ &3 & 25\\

            \hline
        \end{tabular}
        \label{tab:mean-values}
\end{table}

It is worth noting also that we obtained remarkable differences between the GMRT results presented here and those published in \citet{pandey2007}, \citet{godambe2008} and \citet{bhattacharyya2012}. All these differences are related to the \tsys corrections of GMRT data (explained in detail in Appendix~\ref{app:tsys}). We first realised that \citet{godambe2008} and \citet{bhattacharyya2012} did not take into account these corrections, and thus all the flux densities from their data were under-estimated. In fact, the values we obtain before applying the \tsys corrections are compatible with the published ones in these two papers. In the case of \citet{pandey2007}, the origin of the discrepancy is different at each frequency. At 610~MHz, these authors did not apply the mentioned corrections (assuming that the contribution of the Galactic diffuse emission at such frequency is small, while we have found that it is relevant). At 235~MHz these authors obtained the \tsys corrections using the Haslam approximation, which overestimates the flux density in this case (see  Appendix~\ref{app:tsys} and Table~\ref{tab:tsys-corrections}). These are the reasons why \citet{pandey2007} obtained a negative spectral index while we obtain a positive one at these low frequencies.

\subsection{Quasi-simultaneous spectra}

\begin{figure}
        \includegraphics[width=0.485\textwidth]{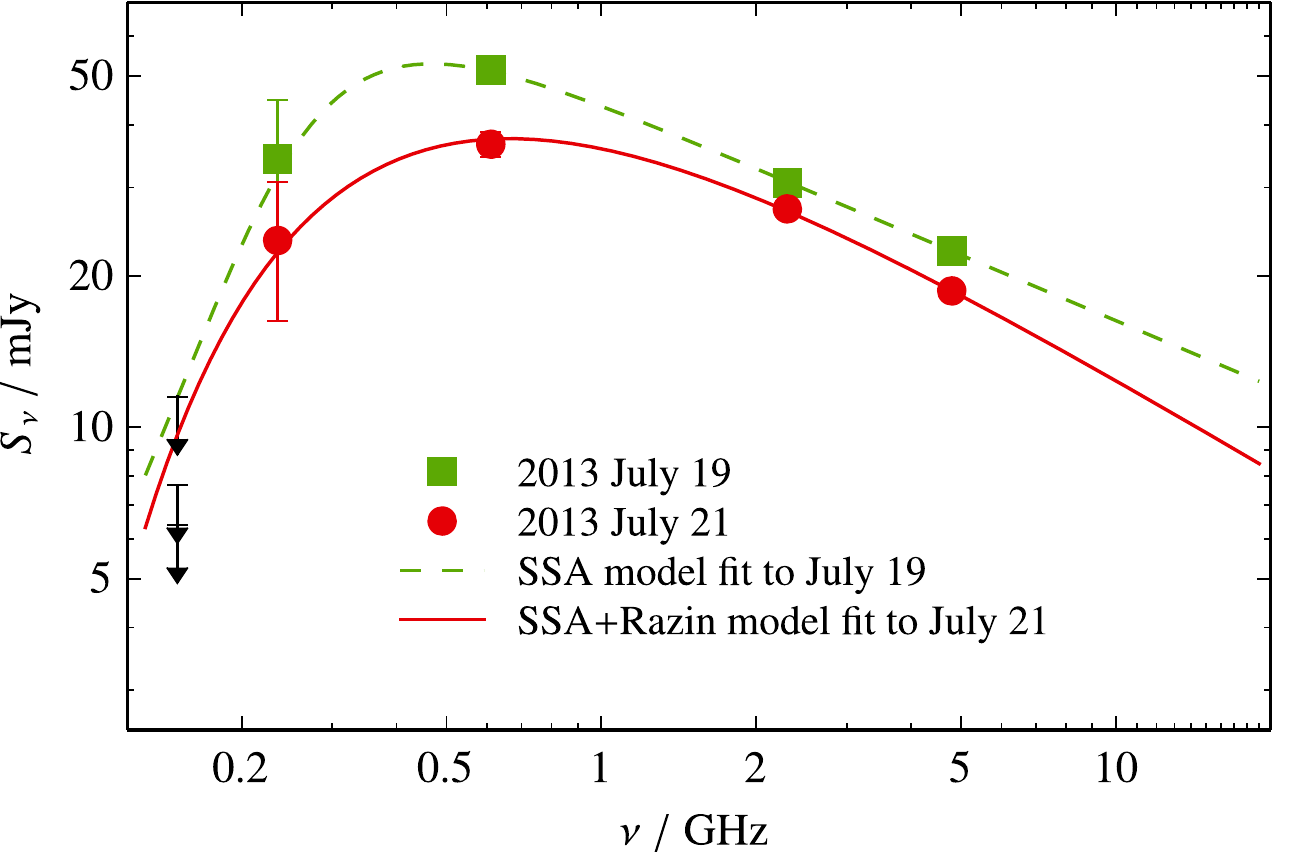}
        \caption{Quasi-simultaneous spectra of LS~5039 obtained with the coordinated GMRT-WSRT campaign in 2013. The data $< 1~\mathrm{GHz}$ were taken with GMRT and the data $> 1~\mathrm{GHz}$ with WSRT. The green squares represent the data of 2013 July 19 ($\phi\approx 0.9$) and the red circles the data of 2013 July 21 ($\phi\approx 0.4$). The black arrows show the 3-$\sigma$ upper limits from the 154~MHz GMRT data of 2013 July 18, 20 and 22. The green dashed line shows a SSA model fit to the July 19 data and the red solid line shows a SSA+Razin model fit to the July 21 data.}
        \label{fig:gmrt-freq-flux}
\end{figure}

To avoid the problem of the non-simultaneity of the data in the previous spectrum, we conducted coordinated GMRT and WSRT observations in 2013 July 19 and 21. The results of these observations are shown in Fig.~\ref{fig:gmrt-freq-flux}. Although the two obtained spectra are qualitatively similar, again with a turnover around $\sim$0.5~GHz, we observe differences between them. The spectrum from 2013 July 19 exhibits a stronger emission and a pure power law behaviour from 0.6 to 5~GHz, while on July 21 the spectrum is slightly curved. The data at 154~MHz were not simultaneous (taken every other day on 2013 July 18, 20 and 22) and provided upper limits on the flux density values of LS~5039. The combination of these three epochs does not improve remarkably the final upper limit, which is similar to the lowest one in Fig.~\ref{fig:gmrt-freq-flux}.


\section{Modeling the LS~5039 spectrum} \label{sec:models}

We will now explain the different models that can fit the observed spectra of LS~5039 in the 0.15--15~GHz range and we will constrain some of the corresponding physical parameters. As discussed previously, the high frequency spectra above 1~GHz can be roughly fitted with a power law with a negative spectral index of $\alpha \approx -0.5$. This, together with the compact radio emission seen at VLBI scales, has led to establishing the synchrotron nature of the radio source \citep{paredes2000}.
In addition, we have also unambiguously revealed the presence of a low-frequency turnover, which takes place at $\sim 0.5~\mathrm{GHz}$.

To explain the observed radio spectrum of LS~5039 we have considered a simple model. The fact that the extended emission is only responsible for a small fraction of the total flux density \citep{moldon2012ls5039} allows us to consider as a first approximation a one-zone model. The absence of a clear orbital modulation in the radio flux density implies that the characteristics of the emitting region must be similar at any orbital phase from the point of view of the observer. The simplest model which verifies this condition is a spherically symmetric emitting region. For simplicity, we have also assumed that the emitting region is isotropic and homogeneous.
In this region we consider the presence of a synchrotron emitting plasma. The turnover below $\sim 0.5~\mathrm{GHz}$ could be produced either by synchrotron self-absorption (SSA), free-free absorption (FFA) or Razin effect. 

Considering a particle injection following a power-law (as expected from synchrotron emission) with index $p$ and normalization $K$ for the mentioned emitting region, the particle density distribution is $n(E)\,\D E = K\,E^{-p}\,\D E$. This injection produces a flux density emission of the form:
\begin{equation}\label{eq:ssa}
S_{\nu} \propto \frac{\Omega\,J(\nu)}{4\pi\,\kappa_{\nu}}\left( 1 - e^{-\kappa_{\nu}\,\ell} \right)
\end{equation}
according to \cite{longair2011}, where $\Omega$ is the solid angle subtended by the source, the emissivity is defined as
\begin{equation}\label{eq:emissivity}
    J(\nu) \propto a(p)\,K\,B^{(p+1)/2}\,\nu^{-(p-1)/2}
\end{equation}
where $a(p)$ is a function that only depends on the index $p$, $B$ is the magnetic field, $\ell$ is the linear size of the emitting region (or the radius in the spherically symmetric case), and $\kappa_{\nu} = \kappa_{\nu}^{\rm SSA} + \kappa_{\nu}^{\rm FFA}$ is the absorption coefficient, where
\begin{equation}
    \kappa_{\nu}^{\rm SSA} \propto K\,B^{(p+2)/2}\,b(p)\,\nu^{-(p+4)/2}
\end{equation}
is the absorption coefficient from the SSA \citep{longair2011}, where $b(p)$ is a function of $p$, and
\begin{equation}
    \kappa_{\nu}^{\rm FFA} \propto n_\mathrm{e}^2\,T_{\mathrm w}^{-3/2}\,\nu^{-2}
\end{equation}
is the absorption coefficient from the FFA \citep{rybicki1979}, where $n_\mathrm{e}$ is the electron number density of the stellar wind and $T_{\mathrm w}$ the wind temperature.
Note that with these definitions $\Omega =\pi\,(\ell/d)^2$, where $d$ is the distance to the source, and $p = 1-2\alpha$, where $\alpha$ is the spectral index.

As the relativistic synchrotron emitting plasma is surrounded by a non-relativistic plasma from the stellar wind of the massive companion, a suppression of the beaming effect from the synchrotron radiation is expected (the mentioned Razin effect). In this case, the emission decreases exponentially at frequencies below a cutoff frequency, $\nu_{\rm R} \equiv 20\,n_\mathrm{e}\,B^{-1}$. This absorption can be approximated by an exponential reduction in the synchrotron emission of Eq.~(\ref{eq:ssa}) with a multiplying factor of $e^{-\nu_{\rm R}/\nu}$ \citep{dougherty2003}.

Different combinations of the mentioned absorption mechanisms have been considered to explain the observed spectra of LS~5039, namely: SSA, FFA, SSA+FFA, SSA+Razin, FFA+Razin, and SSA+FFA+Razin. We have fitted the data and obtained $\chi^2$ statistics from the residuals, from which we have computed the reduced value using the number of degrees of freedom (d.o.f.): $\chi_\mathrm{r}^2 \equiv \chi^2/\mathrm{d.o.f}$.

In the case of the non-simultaneous average spectrum we have taken into account all the data except the values at 2.3~GHz (given that we only have two observations at such frequency, the inferred mean value is not representative enough) and except the upper-limits at 154~MHz. In this case we observe that any of the considered models fits the data with $\chi_\mathrm{r}^2 \ll 1$, implying that their uncertainties are overestimated.
The reason is that these uncertainties reflect the variability of the source, and thus the dispersion of the data, but not the uncertainty of the mean value at each frequency.
Therefore, we need to consider the mean square errors, dividing the standard deviation at a given frequency by the square root of the number of measurements at such frequency.
When using the mean square errors, we observe that there are three models which can explain the spectrum: FFA, SSA and SSA+Razin (see Fig.~\ref{fig:all-freq-flux}-bottom, where mean square errors, within the size of the symbols, are not plotted). The $\chi_\mathrm{r}^2$ from these fits are 2.5, 4.6 and 1.9, respectively. The SSA+Razin model is thus the most accurate model to explain the LS~5039 average emission, and predicts a spectral index of $\alpha_\mathrm{SSA+Razin} = -0.58 \pm 0.02$ for the optically thin part of the spectrum.
Considering the different upper-limits at 154~MHz as possible flux density values at that frequency (i.e., a flux density of $3\sigma\pm\sigma$), we observe that the SSA+Razin is still the best model to reproduce the data (although if the assumed flux density at 154~MHz is as low as $\la7$~mJy, a FFA+Razin seems to be needed).

In the case of the quasi-simultaneous spectra (Fig.~\ref{fig:gmrt-freq-flux}), we have followed the most conservative choice of considering the largest upper-limit at 154~MHz as the possible flux density of LS~5039 at such frequency ($12 \pm 4~\mathrm{mJy}$).
We have fitted the two spectra using all the possible models mentioned above, and assessed their validity with a $\chi^2$ test.
The SSA+Razin model can explain both spectra. However, on 2013 July 19 the contribution of the Razin effect is negligible, and thus we observe a pure SSA.
The FFA model, in contrast, only explains the spectrum on 2013 July 21. The SSA+FFA can explain the two quasi-simultaneous spectra, although with a small contribution of FFA on July 19, but can not explain the average spectrum discussed in the previous paragraph.

In summary, a SSA+Razin model emerges as the most reasonable to explain the spectra of LS~5039, with a different contribution of the Razin effect, which might not be present at all, depending on the epoch.
We note that we only have 1--3 degrees of freedom in the fitting process, implying that in some cases it is subtle to distinguish between models producing similar spectra.

For the discussed models, the underlying physical parameters are coupled. Therefore, the number of free parameters is lower than the number of the physical ones. For the SSA model, we only have three parameters to fit: $p$, $P_1 \equiv \Omega\, B^{-1/2}$, and $P_2 \equiv K\,\ell\,B^{\frac{1}{2}\left(p+2\right)}$. For the SSA+Razin effect model, we have the three previous parameters and an additional one: $\nu_{\rm R} \equiv 20\,n_\mathrm{e}\,B^{-1}$. We can not decouple these physical parameters without making assumptions on some of them. Therefore, we prefer to work directly with the coupled parameters for the different spectra, which are quoted in Table~\ref{tab:fits}. 

\begin{table}
\centering
\caption{Parameters fitted to the averaged data (A.S.) with a SSA+Razin model. The data on 2013 July 19 (J19) were fitted with a SSA model and the data on 2013 July 21 (J21) with a SSA+Razin model. The parameters are: $p \equiv 1-2\,\alpha$, $P_1 \equiv \Omega\, B^{-1/2}$, $P_2 \equiv K\,\ell\,B^{(p+2)/2}$, and $\nu_{\rm R} \equiv 20\,n_\mathrm{e}\,B^{-1}$.}
\label{tab:fits}
\begin{tabular}{l@{~~}c@{~~~}c@{~~~}c@{~~~}c}
	\hline
	Fit & $p$ & $P_1$ & $P_2$ & $\nu_{\rm R}$\\[-1pt]
	&& $\left[\mathrm{10^{-16}\,G^{-1/2}}\right]$ & $\left[ 10^{\,3}\,\mathrm{cm\,G^{(p+2)/2}} \right]$ & $\left[10^{\,8}\,\mathrm{Hz}\right]$\\[+4pt]
	\hline
	A.S.      & $2.16\pm0.04$ 	& $500\pm800$    & $3\pm5$      & $4.1\pm0.2$\\
	J19    & $1.867\pm0.014$	& $3.9\pm0.3$    & $(2.1\pm0.9)\times10^{6}$ & --\\
	J21    & $2.24\pm0.08$ 	& $200\pm600$	 & $0.4\pm1.7$  & $4.1\pm0.7$\\
	\hline
\end{tabular}
\end{table}

\section{Discussion}\label{sec:discussion}

First we will discuss on the physical implications of the modeling of the obtained radio spectra for LS~5039. Second we will discuss on the radio variability of the source, making a comparison with its multi-wavelength emission and the one observed in other gamma-ray binaries.

\subsection{Physical properties from the spectral modeling}

In this work we have reported multi-frequency data of LS~5039 that has allowed us to obtain a non-simultaneous average spectrum in the 0.15--15~GHz range, and two quasi-simultaneous spectra in the 0.15--5~GHz range, showing in all the cases a turnover at $\sim 0.5~\mathrm{GHz}$.
The high-frequency emission is roughly explained by a power-law with a spectral index of $\alpha\approx-0.5$ ($p \sim 2$), which is typical of non-thermal synchrotron radiation, although some indications of curvature below 5~GHz are observed.
All the observed spectra of LS~5039 can be explained with the models which are summarised in Table~\ref{tab:fits}. A SSA+Razin model explains the average spectrum and the 2013 July 21 spectrum, showing both similar properties (on the parameters $p \sim 2.2$ and $\nu_\mathrm{R} \approx 0.41~\mathrm{GHz}$), whereas on 2013 July 19, the spectrum is explained by a pure SSA model, with a negligible contribution of the Razin effect ($\nu_\mathrm{R} \approx 0$). In this case we observe a significant lower index $p = 1.867 \pm 0.014$, a 7-$\sigma$ deviation with respect the average spectrum. The $P_1$ and $P_2$ parameters can only be constrained for the 2013 July 19 spectrum  (see Table~\ref{tab:fits}). Therefore, only in the case of $p$ and $\nu_{\rm R}$ we can claim for changes between the spectra obtained at different epochs. It seems that changes in the injection index but also in the electron density or the magnetic field are present in LS~5039 on timescales of about days.

The Razin effect is widely observed in colliding wind binaries (CWB), where there is a non-relativistic shock between the winds of two massive stars, with typical values of $\nu_{\rm R} \sim 2~\mathrm{GHz}$ \citep{vanloo2005thesis}. Interestingly, the $\nu_{\rm R}$ value inferred for LS~5039 on the average spectrum and on 2013 July 21 is only a factor 5 smaller. The presence of the Razin effect in this gamma-ray binary, and the fact that shocks are present in CWBs, provides further support to the scenario of the young non-accreting pulsar, where shocks also take place, for LS~5039.
We remark that the presence of the Razin effect is also supported by the curvature reported on most of the LS~5039 spectra below 5~GHz, which is easily explained with the consideration of this effect but hardly explained by a SSA or FFA model.

\begin{figure}
        \includegraphics[width=0.48\textwidth]{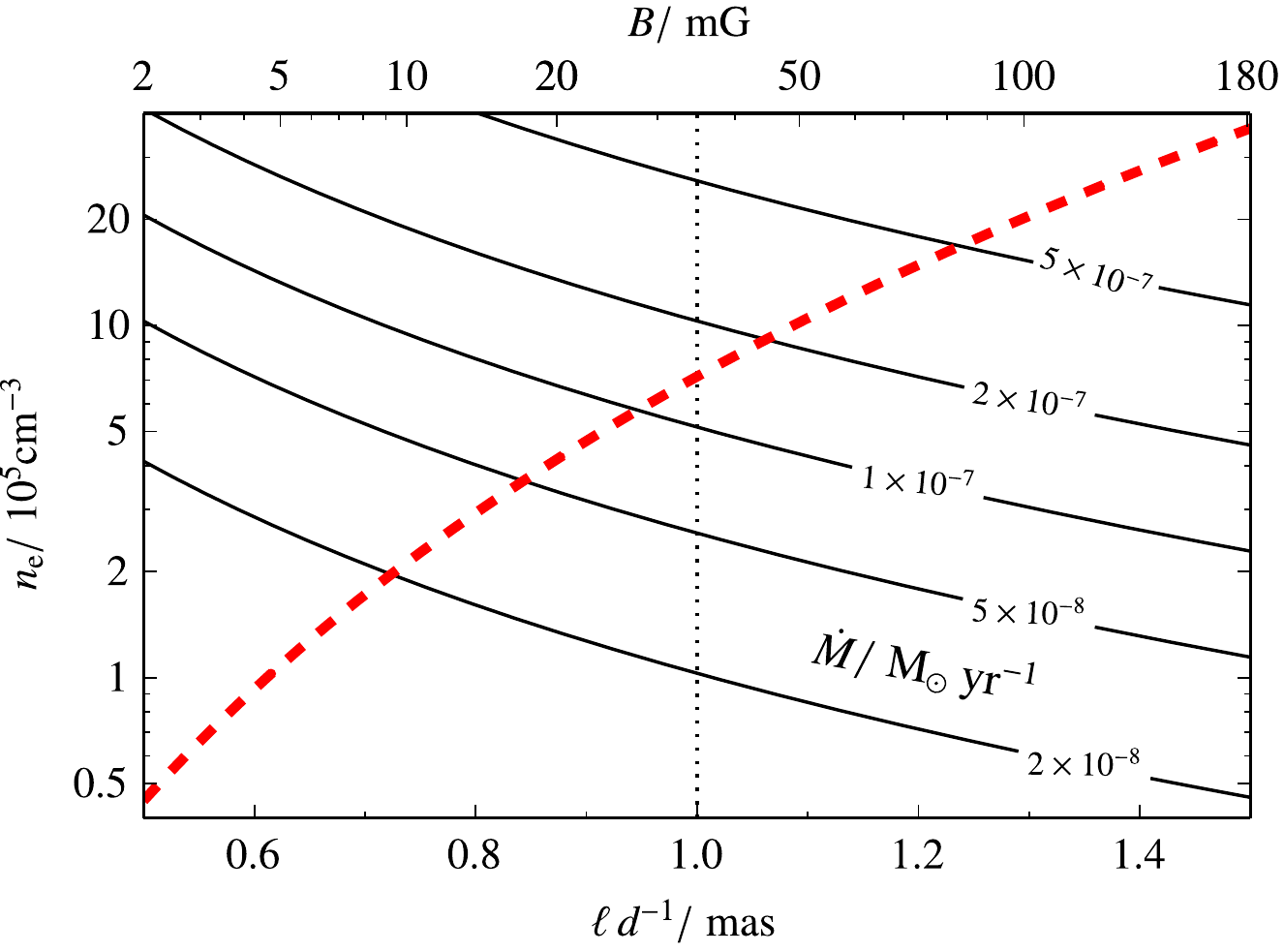}
        \caption{Electron number density for the non-relativistic plasma ($n_\mathrm{e}$) as a function of the angular radius of the emitting region ($\ell\, d^{-1}$, bottom axis) and of the magnetic field value ($B$, top axis). $B$ is determined from the angular radius and the $P_1$ parameter of the pure SSA fit to the 2013 July 19 data (see Table~\ref{tab:fits}). The red dashed line denotes the $n_\mathrm{e}$ values inferred from the $\nu_\mathrm{R}$ parameter of the SSA+Razin fits (which is directly related to $B$). Solid black lines represent the $n_\mathrm{e}$ values obtained with different mass-loss rates $\dot M$. Values below the red dashed line imply an unrealistic mixing above 100~\% of the non-relativistic wind inside the synchrotron radio emitting relativistic plasma. The vertical dotted line shows the central value of 1~mas.}
        \label{fig:mdot}
\end{figure}

The magnetic field $B$ can be estimated from the $P_1$ parameter obtained on the 2013 July 19 spectrum, by assuming a solid angle $\Omega$ for the emitting region, with a radius $\ell$ in the spherically symmetric case considered here.
From \citet{moldon2012ls5039} it is known that most of the radio emission arises from the compact core, which is not resolved with VLBI observations at 5~GHz, and thus has an angular size with a radius $\la 1~\mathrm{mas}$. At the $\sim 3~\mathrm{kpc}$ distance of the source this represents $\sim 3~\mathrm{AU}$, or about 10 times the semimajor axis of the binary system. We consider a solid upper-limit for the angular radius ($\ell\,d^{-1}$) of $\sim 1.5~\mathrm{mas}$, and a lower-limit of $\sim 0.5~\mathrm{mas}$ taking into account the orbital parameters and the absence of orbital modulation (which would be produced by the changing shock geometry and by absorption effects).
From the central value of $1~\mathrm{mas}$ we estimate $B \approx 35~\mathrm{mG}$.
For angular radius in the range of 0.5--1.5~mas we obtain a wide range of magnetic field values of 2--180~mG (see x-axes in Fig.~\ref{fig:mdot}).
These values range from much smaller to close to the $B\sim 200~\mathrm{mG}$ value inferred from the VLBI images at 5~GHz assuming equipartition between the relativistic electrons and the magnetic field \citep{paredes2000}. They are also encompassed with the 3--30~mG range quoted in \citet{bosch-ramon2009ls5039} using luminosity arguments for the lower limit and imposing FFA to derive the upper bound.
The obtained magnetic field value of $B \approx 35~\mathrm{mG}$ for 1~mas allows us to constrain the Lorentz factor of the electrons emitted at $0.5~\mathrm{GHz}$ to $\gamma\sim 60$, confirming that we are in the relativistic regime \citep{pacholczyk1970}.
Using the $\nu_\mathrm{R}$ value obtained from the SSA+Razin fits we derive the electron density for the non-relativistic plasma, $n_\mathrm{e}$, as a function of $B$ (red dashed line in Fig.~\ref{fig:mdot}), with values in the range of $4\times10^4$--$4\times10^6~\mathrm{cm^{-3}}$ for the 0.5--1.5~mas range, and of $7\times10^5~\mathrm{cm^{-3}}$ for the central value of 1~mas.
We can compare these values with the ones obtained from the stellar wind velocity and mass-loss rate. The velocity of the stellar wind is $v_\mathrm{w}\approx 2440~{\rm km\ s^{-1}}$ \citep{mcswain2004}.
Considering a mass-loss rate of $\dot{M} \approx 5 \times 10^{-7}~\mathrm{M_{\sun}\ yr^{-1}}$ \citep{casares2005} we estimate $n_\mathrm{e} \approx 2.6 \times10^6~\mathrm{cm^{-3}}$ for 1~mas, and thus a mixing of $\sim 25~\%$ of the non-relativistic wind inside the synchrotron radio emitting relativistic plasma.

\begin{figure}
        \includegraphics[width=0.48\textwidth]{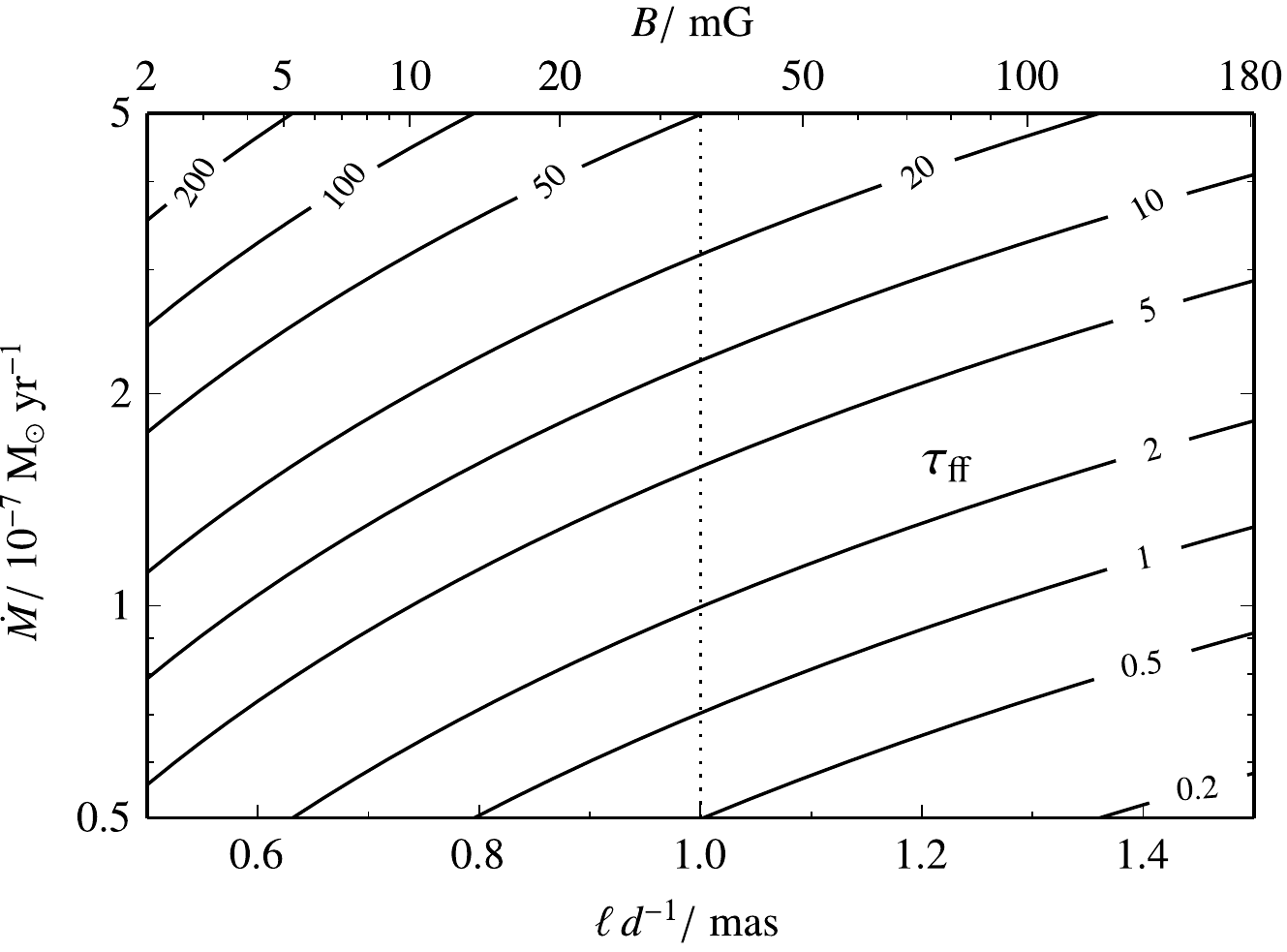}
        \caption{Mass-loss rate ($\dot M$) for different free-free opacities ($\tau_\mathrm{ff}$) as a function of the angular radius of the emitting region ($\ell\, d^{-1}$, bottom axis) and of the magnetic field value ($B$, top axis). The relation between $\ell\, d^{-1}$ and $B$ is the same as in Fig.~\ref{fig:mdot}. Only the region below $\tau_\mathrm{ff}\sim1$ should be considered according to the observations, implying a low value for $\dot M$. The vertical dotted line shows the central value of 1~mas.}
        \label{fig:tau}
\end{figure}

All these values allow us to estimate the free-free opacity, $\tau_\mathrm{ff} \propto \dot M^2\,\nu^{-2}\,\ell^{-3}\,v_\mathrm{w}^{-2}\,T_\mathrm{w}^{-3/2}$ \citep{rybicki1979}. Assuming a temperature for the stellar wind of $T_{\rm w}\approx1.3 \times 10^4~\mathrm{K}$ \citep{krticka2001} we obtain $\tau_{\rm ff} \approx 50$ for an angular radius of 1~mas (see Fig.~\ref{fig:tau}), which is not compatible with the presence of radio emission.
However, from optical spectroscopy there are indications of clumping, which might reduce the mass-loss rate by one order of magnitude down to $\dot{M} \approx 5 \times 10^{-8}~\mathrm{M_{\sun}\ yr^{-1}}$ (Casares et al.\, in preparation). This would yield $\tau_{\rm ff} \approx0.5$ (Fig.~\ref{fig:tau}), and thus the region would be optically thin to free-free opacity. This lower value of $\dot M$ would imply $n_\mathrm{e} \approx 2.6 \times 10^5~\mathrm{cm^{-3}}$, which is smaller than the value derived from the SSA+Razin model fits (see Fig.~\ref{fig:mdot}).
For this mass-loss rate, only angular radii of $\approx 0.85~\mathrm{mas}$ or smaller would be supported (or the mixing would be above 100~\%). At the same time, from Fig.~\ref{fig:tau} we observe that for sizes $\la 0.8~\mathrm{mas}$ we obtain $\tau_\mathrm{ff} \ga 1$. Therefore, the inferred value for the radius of the emitting region is $\sim 0.85~\mathrm{mas}$, implying a significant mixing of the non-relativistic wind inside the synchrotron radio emitting relativistic plasma, even close to $\sim 100~\%$. Another explanation in agreement with this value is the production of the radio emission by secondary electrons produced after photon-photon absorption within the stellar wind \citep*{bosch-ramon2008}.

In summary, from the spectral modeling assuming an homogeneous one-zone emitting region we derive an angular radius of $\sim 0.85~\mathrm{mas}$ ($\ell \sim 2.5~\mathrm{AU}$ at 3~kpc), yielding $B \sim 20~\mathrm{mG}$, and $n_\mathrm{e} \sim 4 \times 10^5~\mathrm{cm^{-3}}$. The mass-loss rate must be around $\dot M\sim 5 \times 10^{-8}~\mathrm{M_{\sun}\ yr^{-1}}$, i.e. about one order of magnitude smaller than the value reported in \citet{casares2005}, thus supporting the presence of a clumpy stellar wind. A high mixing with the relativistic plasma is supported.

\subsection{Variability}\label{sec:variability}

In this work we have reported day to day variability, trends on week timescales and the absence of orbital variability.
The absence of orbital modulation in the radio emission of LS~5039 puts some constraints on the characteristics of the emitting region or radio core, in the sense that they can not change dramatically along the orbit or that the changes in different physical parameters should compensate each other, always as seen from the point of view of the observer.
In addition, the reported stability along the years of the average radio emission at all frequencies implies that this emitting region must be also stable on these timescales. On the other hand, hydrodynamical instabilities in the shocked material or in the outflow properties could explain more easily the observed variability on timescales of days and the trends observed on timescales of weeks \citep{bosch-ramon2009ls5039}.

\begin{table*}
\caption{Properties of the known gamma-ray binaries and their emission. For each binary system we list the spectral type of the companion, the distance to the system, $d$, the orbital period $P_{\rm orb}$, the eccentricity, $e$, the periastron separation, $a(1-e)$, and the apastron separation, $a(1+e)$. We also quote the general behaviour of each binary system: the flux density range at 5~GHz, the spectral index $\alpha$ around this frequency, and the shape of the light-curve at radio and other wavelengths. P means that the light-curve is orbitally modulated, S that the emission is persistent, and U that the source remains undetected. P$^\ast$ means that the light-curve is orbitally modulated but variations from cycle to cycle are also observed.}
\label{tab:other-binaries}
\def\ts{~~~~~}
\begin{tabular}{l@{\ts}l@{\ts}c@{\ts}c@{\ts}c@{\ts}c@{\ts}c@{\ts}c@{\ts}c@{\ts}c@{\ts}c@{\ts}c@{\ts}c}
	\hline
	Source &  Spectral & $d$ & $P_{\rm orb}$ & $e$ & $a(1-e)$ & $a(1+e)$ & $S_{5\,\mathrm{GHz}}$ & $\alpha$ & \multicolumn{4}{c}{Multiwavelength behaviour}\\
	& Type & [kpc] & [d] && [AU] & [AU] & [mJy] && radio & X-ray & GeV & TeV\\
	\hline
	LS~5039$^{(1)}$			&O6.5\,V	& 2.9	 	&3.9		&0.35	&0.09&0.19	&15--30 	&$\sim-0.5$	& S\hphantom{1} &P\hphantom{1}  & P\hphantom{1} & P\hphantom{1}\\
1FGL~J1018.6--5856$^{(2)}$&O6\,V	& 1.9	 	&16.6 	&--		&--   	&--		&1--7 	&$\sim 0.0$		& P\hphantom{1} &P\hphantom{1}  & P\hphantom{1} & P*\\
LS~I~+61~303$^{(3)}$	&B0\,Ve	& 2.0 	&26.5 	&0.72 	&0.1 &0.7		&20--200	&$-0.5$--$+0.5$		& P*	        &P*             & P*            & P*\\
HESS~J0632+057$^{(4)}$	&B0\,Vpe&$\sim1.4$	&315   	&0.83	&0.40&4.35 	&0.1--0.4 	&$\sim-0.6$	        & P*   	        &P*             & U\hphantom{1} & P*\\
PSR~B1259--63$^{(5)}$ 	&O9.5\,Ve	& 2.3	 	&1237  	&0.87	&0.93&13.4	&1--100 	&$-1.0$--$0.0$	& P*	        &P\hphantom{1}  & P\hphantom{1} & P\hphantom{1}\\

	\hline
\end{tabular}

\medskip{{(1)}~\citet{aharonian2005ls5039}, \citet{casares2005}, \citet{kishishita2009}, \citet{abdo2009}, \citet{casares2012}, \citet{zabalza2013};
	{(2)}~\citet{napoli2011}, \citet{fermi2012}, \citet{an2013} \citet{bordas2013};
	{(3)}~\citet{frail1991},  \citet{paredes1997}, \citet{strickman1998}, \citet{harrison2000}, \citet{gregory2002}, \citet{casares2005}, \citet{aragona2009}, \citet{acciari2011}, \citet{hadasch2012}, \citet{ackermann2013};
	{(4)}~\citet{skilton2009}, \citet*{aragona2010}, \citet{bongiorno2011}, \citet{casares2012}, \citet{bordas2012}, \citet{aliu2014};
	{(5)}~\cite{johnston1994}, \citet*{wang2004}, \citet{negueruela2011}, \citet*{dembska2012}, \citet{chernyakova2014}.
}
\end{table*}

In contrast to the radio emission, LS~5039 shows orbitally modulated flux at other wavelengths. The X-ray light-curve is periodic, exhibiting a maximum around orbital phases 0.7--0.8 (during the inferior conjunction of the compact object). It also exhibits small spikes at $\phi \approx 0.40$ and $0.48$ and a strong spike at $\phi \approx 0.70$, probably originated by geometrical effects. The X-ray emission, including these spikes, presents a long-term stability \citep{kishishita2009}.
The TeV light-curve is similar to the X-ray one, exhibiting a maximum around phase 0.7, a minimum flux at superior conjunction, and a strong spike but in this case at phase $\sim 0.8$ \citep{aharonian2005ls5039}. The two light-curves are correlated, with the TeV variability ascribed to photon-photon pair production and anisotropic inverse Compton scattering and the X-ray variability to adiabatic expansion \citep{takahashi2009}.
The GeV light-curve, also periodic, exhibits the maximum emission around the periastron and superior conjunction ($\phi \sim 0.0$), with the minimum emission during the apastron and inferior conjunction ($\phi \sim 0.6$). Therefore, the GeV emission is almost in anti-phase with the X-ray and TeV emission, and is thought to reflect the change in the angle-dependent cross-section of the inverse Compton scattering process and in the photon density \citep{abdo2009}.

The absence of orbital modulation in radio is thus a particular case in the multiwavelength emission of LS~5039.
However, it is interesting to note the deviations from the general behaviour observed around orbital phase 0.8 (as it happens at X-ray and TeV energies). Although in most cases they are not significant enough, we observe an increasing flux density at high frequencies around phase 0.8 (Fig.~\ref{fig:vla-phase-flux}), a higher flux density emission at 610~MHz starting just before phase 0.8 and extending until phase 0.2 (Fig.~\ref{fig:gmrt-phase-flux}), and a large dispersion in the high frequency spectral index close to phase 0.8 (Fig.~\ref{fig:vla-freq-flux}). Since the radio emission of low energy electrons might arise at scales significantly larger than the orbital system size, the fact that we see hints of variability during the inferior conjunction of the compact object at orbital phase 0.8 might suggest the presence of an additional component due to Doppler boosting or to a decrease in the absorption at radio frequencies when the cometary tail is pointing closer to the observer (as invoked by \citealt*{dhawan2006} to explain the radio outbursts of LS~I~+61~303).

To compare the behaviour of LS~5039 with the one detected in the other known gamma-ray binaries, we summarise some of their properties in Table~\ref{tab:other-binaries}. Given the reduced number of known sources and the large differences in their physical properties, a statistical comparison between them is impossible. However, it is remarkable that LS~5039 is the only case in which the radio emission is not orbitally modulated in a periodic or almost periodic way. At high energies (from X-rays to TeV) all of them, including LS~5039, show a periodic emission. That makes even more special the behaviour of the radio emission of this source with respect to the rest of gamma-ray binaries. This behaviour could be related to the fact that LS~5039 presents the shortest orbital period (3.9~d) and the smallest known eccentricity (0.35), which naturally imply a higher absorption of the inner part of the radio outflow.
1FGL~J1018.6$-$5856 can probably be considered as the physically most similar system to LS~5039, because both have similar massive stars (spectral type O6\,V versus O6.5\,V) and the shortest orbital periods (16.6~d versus 3.9~d).
Unfortunately, the eccentricity of 1FGL~J1018.6$-$5856 is still unknown, and in case of being high it could easily explain the observed orbitally modulated radio emission (e.g. due to changes in the absorption processes), contrary to the case of LS~5039.

\section{Summary and conclusions}\label{sec:conclusions}

We have presented a coherent picture of the 0.15--15~GHz spectrum of LS~5039, solving the discrepancies of the low frequency data reported in previous publications. We have unambiguously revealed the presence of a curvature in most of the spectra below 5~GHz and a persistent turnover that takes place at $\sim 0.5~\mathrm{GHz}$. As described in this work, the average spectrum of LS~5039 can be approximated by a simple model with one-zone emitting region, which can be considered homogeneous and spherically symmetric, radiating according to a synchrotron self-absorption model and exhibiting evidences of Razin effect. The Razin effect, reported for first time in a gamma-ray binary, explains the mentioned curvature below 5~GHz. As this effect is commonly observed in colliding wind binaries, it is expected to be present in case that the emission from gamma-ray binaries arises also from a shock originated by the winds of the companion star and the compact object, as it happens in the young non-accreting pulsar scenario.
We observe a certain stability for the parameters of the fits that are well constrained between the average spectrum and particular epochs. However, at other epochs the source shows a slightly different behaviour that implies changes in the injection and in the contribution of the Razin effect (and thus changes in the electron density or the magnetic field).
The presence of free-free absorption can not be discarded in some of the spectra, although the SSA+Razin model is the only one which can explain all the observed spectra.
For angular radii in the range of 0.5--1.5~mas we derive magnetic field values of $B\sim 2$--$180~\mathrm{mG}$. A coherent picture within the one-zone modeling, considering reasonable values of free-free opacity, is obtained for an angular radius of 0.85~mas, $B \sim 20$~mG, $n_\mathrm{e} \sim 4 \times 10^5~\mathrm{cm^{-3}}$, and $\dot M \sim 5 \times 10^{-8}~\mathrm{M_{\sun}\ yr^{-1}}$. These values imply a significant mixing of the stellar wind within the relativistic plasma of the radio outflow. This is the first time that a coherent picture of the physical properties of the assumed one-zone emitting region is presented for LS~5039, including the magnetic field value of the radio emitting plasma.

We have also shown that the radio emission of LS~5039 is persistent with a small variability on day, week and year timescales. This variability is present in all the explored frequency range (0.15--15~GHz), and the relative variation in the flux densities is similar at all frequencies, exhibiting a standard deviation of $\sim 10$--$25~\%$. The absence of orbitally modulated variability constrains the characteristics of the radio emitting region. The observed variability would be produced by stochastic instabilities in the particle injection or in the shocked material, but not by geometrical effects due to the orbital motion. Although the persistence of the flux density was known at 1-yr timescales, we have extend this knowledge up to scales of $\sim 15~\mathrm{yr}$.
At orbital phases $\sim 0.8$ we have detected signatures of an increasing flux density trend at high frequencies, of the starting of an enhanced flux density emission at 610~MHz, and of a large dispersion in the high frequency spectra. Although these signatures are not significant, it is notable that they all happen at the same orbital phase when there is enhanced X-ray and TeV emission. Additional monitoring campaigns, specially at low frequencies but also at high frequencies could clarify if this behaviour is the result of a reduced number of observations or if it is an intrinsic effect from the source due to Doppler boosting or changes in the absorption mechanisms possibly connected with the X-ray or TeV emission.

LS~5039 remains undetected at 154~MHz with the cumulative of $\sim 17~\mathrm{hr}$ of GMRT observations, although according to the modeling its flux density should be close to our upper-limits. A detection with more sensitive interferometric observations would improve our knowledge about the absorbed part of the spectrum. We expect that the higher sensitivity of the Low Frequency Array (LOFAR) would allow to clearly detect the source for first time at such frequency. In the future, other facilities such us LWA, MWA, SKA-low, together with LOFAR, GMRT and the new low-frequency receivers at the VLA, will significantly improve the sensitivity and resolution of low frequency radio observations in the 10~MHz--1~GHz range, thus allowing for detailed studies of absorption in gamma-ray binaries.

\section*{Acknowledgments}

We thank V.~Bosch-Ramon and G.~Romero for helpful discussions on the modeling of the observed radio emission. We thank J.~C.~A.~Miller-Jones for sharing his results on the data analysis of the VLA observations at 330~MHz (project code AM877).
We thank an anonymous referee for providing comments that improved the clarity of the paper.
We thank the staff of the VLA, GMRT and WSRT who made these observations possible.
The Very Long Array is operated by the USA National Radio Astronomy Observatory, which is a facility of the USA National Science Foundation operated under co-operative agreement by Associated Universities, Inc.
GMRT is run by the National Centre for Radio Astrophysics of the Tata Institute of Fundamental Research.
The Westerbork Synthesis Radio Telescope is operated by the ASTRON (Netherlands Institute for Radio Astronomy) with support from the Netherlands Foundation for Scientific Research (NWO).
B.M., M.R. and J.M.P. acknowledge support by the Spanish Ministerio de Econom\'ia y Competitividad (MINECO) under grants AYA2013-47447-C3-1-P and FPA2013-48381-C6-6-P.
B.M. acknowledges financial support from MINECO under grant BES-2011-049886.
J.M.P. acknowledges financial support from ICREA Academia.

\bibliographystyle{mn2e}
\bibliography{bibliography.bib}

\begin{thebibliography}{}
\makeatletter
\relax
\def\mn@urlcharsother{\let\do\@makeother \do\$\do\&\do\#\do\^\do\_\do\%\do\~}
\def\mn@doi{\begingroup\mn@urlcharsother \@ifnextchar [ {\mn@doi@}
  {\mn@doi@[]}}
\def\mn@doi@[#1]#2{\def\@tempa{#1}\ifx\@tempa\@empty \href
  {http://dx.doi.org/#2} {doi:#2}\else \href {http://dx.doi.org/#2} {#1}\fi
  \endgroup}
\def\mn@eprint#1#2{\mn@eprint@#1:#2::\@nil}
\def\mn@eprint@arXiv#1{\href {http://arxiv.org/abs/#1} {{\tt arXiv:#1}}}
\def\mn@eprint@dblp#1{\href {http://dblp.uni-trier.de/rec/bibtex/#1.xml}
  {dblp:#1}}
\def\mn@eprint@#1:#2:#3:#4\@nil{\def\@tempa {#1}\def\@tempb {#2}\def\@tempc
  {#3}\ifx \@tempc \@empty \let \@tempc \@tempb \let \@tempb \@tempa \fi \ifx
  \@tempb \@empty \def\@tempb {arXiv}\fi \@ifundefined
  {mn@eprint@\@tempb}{\@tempb:\@tempc}{\expandafter \expandafter \csname
  mn@eprint@\@tempb\endcsname \expandafter{\@tempc}}}

\bibitem[\protect\citeauthoryear{{Abdo} et~al.,}{{Abdo}
  et~al.}{2009}]{abdo2009}
{Abdo} A.~A.,  et~al., 2009, \mn@doi [\apjl] {10.1088/0004-637X/706/1/L56},
  \href {http://adsabs.harvard.edu/abs/2009ApJ...706L..56A} {706, L56}

\bibitem[\protect\citeauthoryear{{Acciari} et~al.,}{{Acciari}
  et~al.}{2011}]{acciari2011}
{Acciari} V.~A.,  et~al., 2011, \mn@doi [\apj] {10.1088/0004-637X/738/1/3},
  \href {http://adsabs.harvard.edu/abs/2011ApJ...738....3A} {738, 3}

\bibitem[\protect\citeauthoryear{{Ackermann} et~al.,}{{Ackermann}
  et~al.}{2013}]{ackermann2013}
{Ackermann} M.,  et~al., 2013, \mn@doi [\apjl] {10.1088/2041-8205/773/2/L35},
  \href {http://adsabs.harvard.edu/abs/2013ApJ...773L..35A} {773, L35}

\bibitem[\protect\citeauthoryear{{Aharonian} et~al.,}{{Aharonian}
  et~al.}{2005a}]{aharonian2005ls5039}
{Aharonian} F.,  et~al., 2005a, \mn@doi [Science] {10.1126/science.1113764},
  \href {http://adsabs.harvard.edu/abs/2005Sci...309..746A} {309, 746}

\bibitem[\protect\citeauthoryear{{Aharonian} et~al.,}{{Aharonian}
  et~al.}{2005b}]{aharonian2005psr}
{Aharonian} F.,  et~al., 2005b, \mn@doi [\aap] {10.1051/0004-6361:20052983},
  \href {http://adsabs.harvard.edu/abs/2005A%26A...442....1A} {442, 1}

\bibitem[\protect\citeauthoryear{{Aharonian} et~al.,}{{Aharonian}
  et~al.}{2006}]{aharonian2006}
{Aharonian} F.,  et~al., 2006, \mn@doi [\aap] {10.1051/0004-6361:20065940},
  \href {http://adsabs.harvard.edu/abs/2006A%26A...460..743A} {460, 743}

\bibitem[\protect\citeauthoryear{{Albert} et~al.,}{{Albert}
  et~al.}{2006}]{albert2006}
{Albert} J.,  et~al., 2006, \mn@doi [Science] {10.1126/science.1128177}, \href
  {http://adsabs.harvard.edu/abs/2006Sci...312.1771A} {312, 1771}

\bibitem[\protect\citeauthoryear{{Aliu} et~al.,}{{Aliu}
  et~al.}{2014}]{aliu2014}
{Aliu} E.,  et~al., 2014, \mn@doi [\apj] {10.1088/0004-637X/780/2/168}, \href
  {http://adsabs.harvard.edu/abs/2014ApJ...780..168A} {780, 168}

\bibitem[\protect\citeauthoryear{{An}, {Dufour}, {Kaspi}  \& {Harrison}}{{An}
  et~al.}{2013}]{an2013}
{An} H.,  {Dufour} F.,  {Kaspi} V.~M.,   {Harrison} F.~A.,  2013, \mn@doi
  [\apj] {10.1088/0004-637X/775/2/135}, \href
  {http://adsabs.harvard.edu/abs/2013ApJ...775..135A} {775, 135}

\bibitem[\protect\citeauthoryear{{Aragona}, {McSwain}, {Grundstrom}, {Marsh},
  {Roettenbacher}, {Hessler}, {Boyajian}  \& {Ray}}{{Aragona}
  et~al.}{2009}]{aragona2009}
{Aragona} C.,  {McSwain} M.~V.,  {Grundstrom} E.~D.,  {Marsh} A.~N.,
  {Roettenbacher} R.~M.,  {Hessler} K.~M.,  {Boyajian} T.~S.,   {Ray} P.~S.,
  2009, \mn@doi [\apj] {10.1088/0004-637X/698/1/514}, \href
  {http://adsabs.harvard.edu/abs/2009ApJ...698..514A} {698, 514}

\bibitem[\protect\citeauthoryear{{Aragona}, {McSwain}  \& {De
  Becker}}{{Aragona} et~al.}{2010}]{aragona2010}
{Aragona} C.,  {McSwain} M.~V.,   {De Becker} M.,  2010, \mn@doi [\apj]
  {10.1088/0004-637X/724/1/306}, \href
  {http://adsabs.harvard.edu/abs/2010ApJ...724..306A} {724, 306}

\bibitem[\protect\citeauthoryear{{Bhattacharyya}, {Godambe}, {Bhatt}, {Mitra}
  \& {Choudhury}}{{Bhattacharyya} et~al.}{2012}]{bhattacharyya2012}
{Bhattacharyya} S.,  {Godambe} S.,  {Bhatt} N.,  {Mitra} A.,   {Choudhury} M.,
  2012, \mn@doi [\mnras] {10.1111/j.1745-3933.2011.01190.x}, \href
  {http://adsabs.harvard.edu/abs/2012MNRAS.421L...1B} {421, L1}

\bibitem[\protect\citeauthoryear{{Bongiorno}, {Falcone}, {Stroh}, {Holder},
  {Skilton}, {Hinton}, {Gehrels}  \& {Grube}}{{Bongiorno}
  et~al.}{2011}]{bongiorno2011}
{Bongiorno} S.~D.,  {Falcone} A.~D.,  {Stroh} M.,  {Holder} J.,  {Skilton}
  J.~L.,  {Hinton} J.~A.,  {Gehrels} N.,   {Grube} J.,  2011, \mn@doi [\apjl]
  {10.1088/2041-8205/737/1/L11}, \href
  {http://adsabs.harvard.edu/abs/2011ApJ...737L..11B} {737, L11}

\bibitem[\protect\citeauthoryear{{Bordas}, {H.~E.~S.~S.~Collaboration}, {Maier}
   \& {VERITAS Collaboration}}{{Bordas} et~al.}{2012}]{bordas2012}
{Bordas} P.,  {H.~E.~S.~S.~Collaboration} {Maier} G.,   {VERITAS Collaboration}
  2012, in {Aharonian} F.~A.,  {Hofmann} W.,   {Rieger} F.~M.,  eds,  AIP
  Conference Series Vol. 1505, HIGH ENERGY GAMMA-RAY ASTRONOMY: 5th
  International Meeting on High Energy Gamma-Ray Astronomy. p.~366

\bibitem[\protect\citeauthoryear{{Bordas} et~al.,}{{Bordas}
  et~al.}{2013}]{bordas2013}
{Bordas} P.,  et~al., 2013, in Proceedings of the 33rd International Cosmic Ray
  Conference (ICRC2013).  (\mn@eprint {arXiv} {1307.6262})

\bibitem[\protect\citeauthoryear{{Bosch-Ramon}}{{Bosch-Ramon}}{2009}]{bosch-ramon2009ls5039}
{Bosch-Ramon} V.,  2009, \mn@doi [\aap] {10.1051/0004-6361:200810969}, \href
  {http://adsabs.harvard.edu/abs/2009A%26A...493..829B} {493, 829}

\bibitem[\protect\citeauthoryear{{Bosch-Ramon}}{{Bosch-Ramon}}{2013}]{bosch-ramon2013}
{Bosch-Ramon} V.,  2013, \mn@doi [\aap] {10.1051/0004-6361/201322249}, \href
  {http://adsabs.harvard.edu/abs/2013A%26A...560A..32B} {560, A32}

\bibitem[\protect\citeauthoryear{{Bosch-Ramon}, {Paredes}, {Rib{\'o}},
  {Miller}, {Reig}  \& {Mart{\'{\i}}}}{{Bosch-Ramon}
  et~al.}{2005}]{bosch-ramon2005}
{Bosch-Ramon} V.,  {Paredes} J.~M.,  {Rib{\'o}} M.,  {Miller} J.~M.,  {Reig}
  P.,   {Mart{\'{\i}}} J.,  2005, \mn@doi [\apj] {10.1086/429901}, \href
  {http://adsabs.harvard.edu/abs/2005ApJ...628..388B} {628, 388}

\bibitem[\protect\citeauthoryear{{Bosch-Ramon}, {Khangulyan}  \&
  {Aharonian}}{{Bosch-Ramon} et~al.}{2008}]{bosch-ramon2008}
{Bosch-Ramon} V.,  {Khangulyan} D.,   {Aharonian} F.~A.,  2008, \mn@doi [\aap]
  {10.1051/0004-6361:20079252}, \href
  {http://adsabs.harvard.edu/abs/2008A%26A...482..397B} {482, 397}

\bibitem[\protect\citeauthoryear{{Casares}, {Rib{\'o}}, {Ribas}, {Paredes},
  {Mart{\'{\i}}}  \& {Herrero}}{{Casares} et~al.}{2005}]{casares2005}
{Casares} J.,  {Rib{\'o}} M.,  {Ribas} I.,  {Paredes} J.~M.,  {Mart{\'{\i}}}
  J.,   {Herrero} A.,  2005, \mn@doi [\mnras]
  {10.1111/j.1365-2966.2005.09617.x}, \href
  {http://adsabs.harvard.edu/abs/2005MNRAS.364..899C} {364, 899}

\bibitem[\protect\citeauthoryear{{Casares}, {Rib{\'o}}, {Ribas}, {Paredes},
  {Vilardell}  \& {Negueruela}}{{Casares} et~al.}{2012}]{casares2012}
{Casares} J.,  {Rib{\'o}} M.,  {Ribas} I.,  {Paredes} J.~M.,  {Vilardell} F.,
  {Negueruela} I.,  2012, \mn@doi [\mnras] {10.1111/j.1365-2966.2011.20368.x},
  \href {http://adsabs.harvard.edu/abs/2012MNRAS.421.1103C} {421, 1103}

\bibitem[\protect\citeauthoryear{{Chernyakova} et~al.,}{{Chernyakova}
  et~al.}{2014}]{chernyakova2014}
{Chernyakova} M.,  et~al., 2014, \mn@doi [\mnras] {10.1093/mnras/stu021}, \href
  {http://adsabs.harvard.edu/abs/2014MNRAS.439..432C} {439, 432}

\bibitem[\protect\citeauthoryear{{Clark} et~al.,}{{Clark}
  et~al.}{2001}]{clark2001}
{Clark} J.~S.,  et~al., 2001, \mn@doi [\aap] {10.1051/0004-6361:20010919},
  \href {http://adsabs.harvard.edu/abs/2001A%26A...376..476C} {376, 476}

\bibitem[\protect\citeauthoryear{{Cotton}}{{Cotton}}{2008}]{cotton2008}
{Cotton} W.~D.,  2008, \mn@doi [\pasp] {10.1086/586754}, \href
  {http://adsabs.harvard.edu/abs/2008PASP..120..439C} {120, 439}

\bibitem[\protect\citeauthoryear{{Crane} \& {Napier}}{{Crane} \&
  {Napier}}{1989}]{crane1989}
{Crane} P.~C.,  {Napier} P.~J.,  1989, in {Perley} R.~A.,  {Schwab} F.~R.,
  {Bridle} A.~H.,  eds,  ASP Conf. Ser. Vol. 6, Synthesis Imaging in Radio
  Astronomy. p.~139

\bibitem[\protect\citeauthoryear{{Dembska}, {Kijak}  \&
  {Lewandowski}}{{Dembska} et~al.}{2012}]{dembska2012}
{Dembska} M.,  {Kijak} J.,   {Lewandowski} W.,  2012, in {Lewandowski} W.,
  {Maron} O.,   {Kijak} J.,  eds,  ASP Conf. Ser. Vol. 466, Electromagnetic
  Radiation from Pulsars and Magnetars. p.~75

\bibitem[\protect\citeauthoryear{{Dhawan}, {Mioduszewski}  \& {Rupen}}{{Dhawan}
  et~al.}{2006}]{dhawan2006}
{Dhawan} V.,  {Mioduszewski} A.,   {Rupen} M.,  2006, in VI Microquasar
  Workshop: Microquasars and Beyond. p.~52

\bibitem[\protect\citeauthoryear{{Dougherty}, {Pittard}, {Kasian}, {Coker},
  {Williams}  \& {Lloyd}}{{Dougherty} et~al.}{2003}]{dougherty2003}
{Dougherty} S.~M.,  {Pittard} J.~M.,  {Kasian} L.,  {Coker} R.~F.,  {Williams}
  P.~M.,   {Lloyd} H.~M.,  2003, \mn@doi [\aap] {10.1051/0004-6361:20031048},
  \href {http://adsabs.harvard.edu/abs/2003A%26A...409..217D} {409, 217}

\bibitem[\protect\citeauthoryear{{Dubus}}{{Dubus}}{2006}]{dubus2006}
{Dubus} G.,  2006, \mn@doi [\aap] {10.1051/0004-6361:20054779}, \href
  {http://adsabs.harvard.edu/abs/2006A%26A...456..801D} {456, 801}

\bibitem[\protect\citeauthoryear{{Dubus}}{{Dubus}}{2013}]{dubus2013}
{Dubus} G.,  2013, \mn@doi [\aapr] {10.1007/s00159-013-0064-5}, \href
  {http://adsabs.harvard.edu/abs/2013A%26ARv..21...64D} {21, 64}

\bibitem[\protect\citeauthoryear{{Durant}, {Kargaltsev}, {Pavlov}, {Chang}  \&
  {Garmire}}{{Durant} et~al.}{2011}]{durant2011}
{Durant} M.,  {Kargaltsev} O.,  {Pavlov} G.~G.,  {Chang} C.,   {Garmire} G.~P.,
   2011, \mn@doi [\apj] {10.1088/0004-637X/735/1/58}, \href
  {http://adsabs.harvard.edu/abs/2011ApJ...735...58D} {735, 58}

\bibitem[\protect\citeauthoryear{{Fermi LAT Collaboration} et~al.,}{{Fermi LAT
  Collaboration} et~al.}{2012}]{fermi2012}
{Fermi LAT Collaboration} et~al., 2012, \mn@doi [Science]
  {10.1126/science.1213974}, \href
  {http://adsabs.harvard.edu/abs/2012Sci...335..189F} {335, 189}

\bibitem[\protect\citeauthoryear{{Frail} \& {Hjellming}}{{Frail} \&
  {Hjellming}}{1991}]{frail1991}
{Frail} D.~A.,  {Hjellming} R.~M.,  1991, \mn@doi [\aj] {10.1086/115833}, \href
  {http://adsabs.harvard.edu/abs/1991AJ....101.2126F} {101, 2126}

\bibitem[\protect\citeauthoryear{{Godambe}, {Bhattacharyya}, {Bhatt}  \&
  {Choudhury}}{{Godambe} et~al.}{2008}]{godambe2008}
{Godambe} S.,  {Bhattacharyya} S.,  {Bhatt} N.,   {Choudhury} M.,  2008,
  \mn@doi [\mnras] {10.1111/j.1745-3933.2008.00532.x}, \href
  {http://adsabs.harvard.edu/abs/2008MNRAS.390L..43G} {390, L43}

\bibitem[\protect\citeauthoryear{{Gregory}}{{Gregory}}{2002}]{gregory2002}
{Gregory} P.~C.,  2002, \mn@doi [\apj] {10.1086/341257}, \href
  {http://adsabs.harvard.edu/abs/2002ApJ...575..427G} {575, 427}

\bibitem[\protect\citeauthoryear{{Hadasch} et~al.,}{{Hadasch}
  et~al.}{2012}]{hadasch2012}
{Hadasch} D.,  et~al., 2012, \mn@doi [\apj] {10.1088/0004-637X/749/1/54}, \href
  {http://adsabs.harvard.edu/abs/2012ApJ...749...54H} {749, 54}

\bibitem[\protect\citeauthoryear{{Harrison}, {Ray}, {Leahy}, {Waltman}  \&
  {Pooley}}{{Harrison} et~al.}{2000}]{harrison2000}
{Harrison} F.~A.,  {Ray} P.~S.,  {Leahy} D.~A.,  {Waltman} E.~B.,   {Pooley}
  G.~G.,  2000, \mn@doi [\apj] {10.1086/308157}, \href
  {http://adsabs.harvard.edu/abs/2000ApJ...528..454H} {528, 454}

\bibitem[\protect\citeauthoryear{{Haslam}, {Salter}, {Stoffel}  \&
  {Wilson}}{{Haslam} et~al.}{1982}]{haslam1982}
{Haslam} C.~G.~T.,  {Salter} C.~J.,  {Stoffel} H.,   {Wilson} W.~E.,  1982,
  \aaps, \href {http://adsabs.harvard.edu/abs/1982A%26AS...47....1H} {47, 1}

\bibitem[\protect\citeauthoryear{{Hinton} et~al.,}{{Hinton}
  et~al.}{2009}]{hinton2009}
{Hinton} J.~A.,  et~al., 2009, \mn@doi [\apjl] {10.1088/0004-637X/690/2/L101},
  \href {http://adsabs.harvard.edu/abs/2009ApJ...690L.101H} {690, L101}

\bibitem[\protect\citeauthoryear{{Hornby} \& {Williams}}{{Hornby} \&
  {Williams}}{1966}]{hornby1966}
{Hornby} J.~M.,  {Williams} P.~J.~S.,  1966, \mnras, \href
  {http://adsabs.harvard.edu/abs/1966MNRAS.131..237H} {131, 237}

\bibitem[\protect\citeauthoryear{{Intema}, {van der Tol}, {Cotton}, {Cohen},
  {van Bemmel}  \& {R{\"o}ttgering}}{{Intema} et~al.}{2009}]{intema2009}
{Intema} H.~T.,  {van der Tol} S.,  {Cotton} W.~D.,  {Cohen} A.~S.,  {van
  Bemmel} I.~M.,   {R{\"o}ttgering} H.~J.~A.,  2009, \mn@doi [\aap]
  {10.1051/0004-6361/200811094}, \href
  {http://adsabs.harvard.edu/abs/2009A%26A...501.1185I} {501, 1185}

\bibitem[\protect\citeauthoryear{{Johnston}, {Manchester}, {Lyne}, {Nicastro}
  \& {Spyromilio}}{{Johnston} et~al.}{1994}]{johnston1994}
{Johnston} S.,  {Manchester} R.~N.,  {Lyne} A.~G.,  {Nicastro} L.,
  {Spyromilio} J.,  1994, \mnras, \href
  {http://adsabs.harvard.edu/abs/1994MNRAS.268..430J} {268, 430}

\bibitem[\protect\citeauthoryear{{Kettenis}, {van Langevelde}, {Reynolds}  \&
  {Cotton}}{{Kettenis} et~al.}{2006}]{kettenis2006}
{Kettenis} M.,  {van Langevelde} H.~J.,  {Reynolds} C.,   {Cotton} B.,  2006,
  in {Gabriel} C.,  {Arviset} C.,  {Ponz} D.,   {Enrique} S.,  eds,  ASP Conf.
  Ser. Vol. 351, Astronomical Data Analysis Software and Systems XV. p.~497

\bibitem[\protect\citeauthoryear{{Kishishita}, {Tanaka}, {Uchiyama}  \&
  {Takahashi}}{{Kishishita} et~al.}{2009}]{kishishita2009}
{Kishishita} T.,  {Tanaka} T.,  {Uchiyama} Y.,   {Takahashi} T.,  2009, \mn@doi
  [\apjl] {10.1088/0004-637X/697/1/L1}, \href
  {http://adsabs.harvard.edu/abs/2009ApJ...697L...1K} {697, L1}

\bibitem[\protect\citeauthoryear{{Krti{\v c}ka} \& {Kub{\'a}t}}{{Krti{\v c}ka}
  \& {Kub{\'a}t}}{2001}]{krticka2001}
{Krti{\v c}ka} J.,  {Kub{\'a}t} J.,  2001, \mn@doi [\aap]
  {10.1051/0004-6361:20011075}, \href
  {http://adsabs.harvard.edu/abs/2001A%26A...377..175K} {377, 175}

\bibitem[\protect\citeauthoryear{{Longair}}{{Longair}}{2011}]{longair2011}
{Longair} M.~S.,  2011, {High Energy Astrophysics}.
Cambridge: Cambridge University Press

\bibitem[\protect\citeauthoryear{{Mart\'i}, {Paredes}  \& {Rib\'o}}{{Mart\'i}
  et~al.}{1998}]{marti1998}
{Mart\'i} J.,  {Paredes} J.~M.,   {Rib\'o} M.,  1998, \aap, \href
  {http://adsabs.harvard.edu/abs/1998A%26A...338L..71M} {338, L71}

\bibitem[\protect\citeauthoryear{{Martocchia}, {Motch}  \&
  {Negueruela}}{{Martocchia} et~al.}{2005}]{martocchia2005}
{Martocchia} A.,  {Motch} C.,   {Negueruela} I.,  2005, \mn@doi [\aap]
  {10.1051/0004-6361:20041390}, \href
  {http://adsabs.harvard.edu/abs/2005A%26A...430..245M} {430, 245}

\bibitem[\protect\citeauthoryear{{McSwain}, {Gies}, {Huang}, {Wiita}, {Wingert}
   \& {Kaper}}{{McSwain} et~al.}{2004}]{mcswain2004}
{McSwain} M.~V.,  {Gies} D.~R.,  {Huang} W.,  {Wiita} P.~J.,  {Wingert} D.~W.,
   {Kaper} L.,  2004, \mn@doi [\apj] {10.1086/379892}, \href
  {http://adsabs.harvard.edu/abs/2004ApJ...600..927M} {600, 927}

\bibitem[\protect\citeauthoryear{{McSwain}, {Ray}, {Ransom}, {Roberts},
  {Dougherty}  \& {Pooley}}{{McSwain} et~al.}{2011}]{mcswain2011}
{McSwain} M.~V.,  {Ray} P.~S.,  {Ransom} S.~M.,  {Roberts} M.~S.~E.,
  {Dougherty} S.~M.,   {Pooley} G.~G.,  2011, \mn@doi [\apj]
  {10.1088/0004-637X/738/1/105}, \href
  {http://adsabs.harvard.edu/abs/2011ApJ...738..105M} {738, 105}

\bibitem[\protect\citeauthoryear{{Mold{\'o}n}, {Rib{\'o}}, {Paredes},
  {Brisken}, {Dhawan}, {Kramer}, {Lyne}  \& {Stappers}}{{Mold{\'o}n}
  et~al.}{2012a}]{moldon2012}
{Mold{\'o}n} J.,  {Rib{\'o}} M.,  {Paredes} J.~M.,  {Brisken} W.,  {Dhawan} V.,
   {Kramer} M.,  {Lyne} A.~G.,   {Stappers} B.~W.,  2012a, \mn@doi [\aap]
  {10.1051/0004-6361/201219205}, \href
  {http://adsabs.harvard.edu/abs/2012A%26A...543A..26M} {543, A26}

\bibitem[\protect\citeauthoryear{{Mold\'on}, {Rib\'o}  \& {Paredes}}{{Mold\'on}
  et~al.}{2012b}]{moldon2012ls5039}
{Mold\'on} J.,  {Rib\'o} M.,   {Paredes} J.~M.,  2012b, \mn@doi [\aap]
  {10.1051/0004-6361/201219553}, \href
  {http://adsabs.harvard.edu/abs/2012A%26A...548A.103M} {548, A103}

\bibitem[\protect\citeauthoryear{{Motch}, {Haberl}, {Dennerl}, {Pakull}  \&
  {Janot-Pacheco}}{{Motch} et~al.}{1997}]{motch1997}
{Motch} C.,  {Haberl} F.,  {Dennerl} K.,  {Pakull} M.,   {Janot-Pacheco} E.,
  1997, \aap, \href {http://adsabs.harvard.edu/abs/1997A%26A...323..853M} {323,
  853}

\bibitem[\protect\citeauthoryear{{Napoli}, {McSwain}, {Boyer}  \&
  {Roettenbacher}}{{Napoli} et~al.}{2011}]{napoli2011}
{Napoli} V.~J.,  {McSwain} M.~V.,  {Boyer} A.~N.~M.,   {Roettenbacher} R.~M.,
  2011, \mn@doi [\pasp] {10.1086/662692}, \href
  {http://adsabs.harvard.edu/abs/2011PASP..123.1262N} {123, 1262}

\bibitem[\protect\citeauthoryear{{Negueruela}, {Rib{\'o}}, {Herrero},
  {Lorenzo}, {Khangulyan}  \& {Aharonian}}{{Negueruela}
  et~al.}{2011}]{negueruela2011}
{Negueruela} I.,  {Rib{\'o}} M.,  {Herrero} A.,  {Lorenzo} J.,  {Khangulyan}
  D.,   {Aharonian} F.~A.,  2011, \mn@doi [\apjl]
  {10.1088/2041-8205/732/1/L11}, \href
  {http://adsabs.harvard.edu/abs/2011ApJ...732L..11N} {732, L11}

\bibitem[\protect\citeauthoryear{{Pacholczyk}}{{Pacholczyk}}{1970}]{pacholczyk1970}
{Pacholczyk} A.~G.,  1970, {Radio astrophysics. Nonthermal processes in
  galactic and extragalactic sources}.
San Francisco: Freeman

\bibitem[\protect\citeauthoryear{{Pandey}, {Rao}, {Ishwara-Chandra},
  {Durouchoux}  \& {Manchanda}}{{Pandey} et~al.}{2007}]{pandey2007}
{Pandey} M.,  {Rao} A.~P.,  {Ishwara-Chandra} C.~H.,  {Durouchoux} P.,
  {Manchanda} R.~K.,  2007, \mn@doi [\aap] {10.1051/0004-6361:20065856}, \href
  {http://adsabs.harvard.edu/abs/2007A%26A...463..567P} {463, 567}

\bibitem[\protect\citeauthoryear{{Paredes}, {Marti}, {Peracaula}  \&
  {Ribo}}{{Paredes} et~al.}{1997}]{paredes1997}
{Paredes} J.~M.,  {Marti} J.,  {Peracaula} M.,   {Ribo} M.,  1997, \aap, \href
  {http://adsabs.harvard.edu/abs/1997A%26A...320L..25P} {320, L25}

\bibitem[\protect\citeauthoryear{{Paredes}, {Mart{\'{\i}}}, {Rib{\'o}}  \&
  {Massi}}{{Paredes} et~al.}{2000}]{paredes2000}
{Paredes} J.~M.,  {Mart{\'{\i}}} J.,  {Rib{\'o}} M.,   {Massi} M.,  2000,
  \mn@doi [Science] {10.1126/science.288.5475.2340}, \href
  {http://adsabs.harvard.edu/abs/2000Sci...288.2340P} {288, 2340}

\bibitem[\protect\citeauthoryear{{Paredes}, {Rib{\'o}}, {Ros}, {Mart{\'{\i}}}
  \& {Massi}}{{Paredes} et~al.}{2002}]{paredes2002}
{Paredes} J.~M.,  {Rib{\'o}} M.,  {Ros} E.,  {Mart{\'{\i}}} J.,   {Massi} M.,
  2002, \mn@doi [\aap] {10.1051/0004-6361:20021256}, \href
  {http://adsabs.harvard.edu/abs/2002A%26A...393L..99P} {393, L99}

\bibitem[\protect\citeauthoryear{{Rea}, {Torres}, {Caliandro}, {Hadasch}, {van
  der Klis}, {Jonker}, {M{\'e}ndez}  \& {Sierpowska-Bartosik}}{{Rea}
  et~al.}{2011}]{rea2011}
{Rea} N.,  {Torres} D.~F.,  {Caliandro} G.~A.,  {Hadasch} D.,  {van der Klis}
  M.,  {Jonker} P.~G.,  {M{\'e}ndez} M.,   {Sierpowska-Bartosik} A.,  2011,
  \mn@doi [\mnras] {10.1111/j.1365-2966.2011.19148.x}, \href
  {http://adsabs.harvard.edu/abs/2011MNRAS.416.1514R} {416, 1514}

\bibitem[\protect\citeauthoryear{{Reig}, {Rib{\'o}}, {Paredes}  \&
  {Mart{\'{\i}}}}{{Reig} et~al.}{2003}]{reig2003}
{Reig} P.,  {Rib{\'o}} M.,  {Paredes} J.~M.,   {Mart{\'{\i}}} J.,  2003,
  \mn@doi [\aap] {10.1051/0004-6361:20030674}, \href
  {http://adsabs.harvard.edu/abs/2003A%26A...405..285R} {405, 285}

\bibitem[\protect\citeauthoryear{{Rib{\'o}}, {Reig}, {Mart{\'{\i}}}  \&
  {Paredes}}{{Rib{\'o}} et~al.}{1999}]{ribo1999}
{Rib{\'o}} M.,  {Reig} P.,  {Mart{\'{\i}}} J.,   {Paredes} J.~M.,  1999, \aap,
  \href {http://adsabs.harvard.edu/abs/1999A%26A...347..518R} {347, 518}

\bibitem[\protect\citeauthoryear{{Rib{\'o}}, {Paredes}, {Mold{\'o}n},
  {Mart{\'{\i}}}  \& {Massi}}{{Rib{\'o}} et~al.}{2008}]{ribo2008}
{Rib{\'o}} M.,  {Paredes} J.~M.,  {Mold{\'o}n} J.,  {Mart{\'{\i}}} J.,
  {Massi} M.,  2008, \mn@doi [\aap] {10.1051/0004-6361:20078390}, \href
  {http://adsabs.harvard.edu/abs/2008A%26A...481...17R} {481, 17}

\bibitem[\protect\citeauthoryear{{Roger}, {Costain}, {Landecker}  \&
  {Swerdlyk}}{{Roger} et~al.}{1999}]{roger1999}
{Roger} R.~S.,  {Costain} C.~H.,  {Landecker} T.~L.,   {Swerdlyk} C.~M.,  1999,
  \mn@doi [\aaps] {10.1051/aas:1999239}, \href
  {http://adsabs.harvard.edu/abs/1999A%26AS..137....7R} {137, 7}

\bibitem[\protect\citeauthoryear{{Rybicki} \& {Lightman}}{{Rybicki} \&
  {Lightman}}{1979}]{rybicki1979}
{Rybicki} G.~B.,  {Lightman} A.~P.,  1979, {Radiative processes in
  astrophysics}.
New York: Wiley-Interscience

\bibitem[\protect\citeauthoryear{{Sirothia}}{{Sirothia}}{2009}]{sirothia2009}
{Sirothia} S.~K.,  2009, \mn@doi [\mnras] {10.1111/j.1365-2966.2009.14993.x},
  \href {http://adsabs.harvard.edu/abs/2009MNRAS.398..853S} {398, 853}

\bibitem[\protect\citeauthoryear{{Skilton} et~al.,}{{Skilton}
  et~al.}{2009}]{skilton2009}
{Skilton} J.~L.,  et~al., 2009, \mn@doi [\mnras]
  {10.1111/j.1365-2966.2009.15272.x}, \href
  {http://adsabs.harvard.edu/abs/2009MNRAS.399..317S} {399, 317}

\bibitem[\protect\citeauthoryear{{Strickman}, {Tavani}, {Coe}, {Steele},
  {Fabregat}, {Marti}, {Paredes}  \& {Ray}}{{Strickman}
  et~al.}{1998}]{strickman1998}
{Strickman} M.~S.,  {Tavani} M.,  {Coe} M.~J.,  {Steele} I.~A.,  {Fabregat} J.,
   {Marti} J.,  {Paredes} J.~M.,   {Ray} P.~S.,  1998, \mn@doi [\apj]
  {10.1086/305446}, \href {http://adsabs.harvard.edu/abs/1998ApJ...497..419S}
  {497, 419}

\bibitem[\protect\citeauthoryear{{Takahashi} et~al.,}{{Takahashi}
  et~al.}{2009}]{takahashi2009}
{Takahashi} T.,  et~al., 2009, \mn@doi [\apj] {10.1088/0004-637X/697/1/592},
  \href {http://adsabs.harvard.edu/abs/2009ApJ...697..592T} {697, 592}

\bibitem[\protect\citeauthoryear{{Thompson}}{{Thompson}}{1999}]{thompson1999}
{Thompson} A.~R.,  1999, in {Taylor} G.~B.,  {Carilli} C.~L.,   {Perley} R.~A.,
   eds,  ASP Conf. Ser. Vol. 180, Synthesis Imaging in Radio Astronomy II.
  p.~11

\bibitem[\protect\citeauthoryear{{Van Loo}}{{Van Loo}}{2005}]{vanloo2005thesis}
{Van Loo} S.,  2005, PhD thesis, Royal Observatory of Belgium Ringlaan 3 B-1180
  Brussels, \url {http://www.ast.leeds.ac.uk/~svenvl/journals/thesis.pdf}

\bibitem[\protect\citeauthoryear{{Wang}, {Johnston}  \& {Manchester}}{{Wang}
  et~al.}{2004}]{wang2004}
{Wang} N.,  {Johnston} S.,   {Manchester} R.~N.,  2004, \mn@doi [\mnras]
  {10.1111/j.1365-2966.2004.07806.x}, \href
  {http://adsabs.harvard.edu/abs/2004MNRAS.351..599W} {351, 599}

\bibitem[\protect\citeauthoryear{{Zabalza}, {Bosch-Ramon}, {Aharonian}  \&
  {Khangulyan}}{{Zabalza} et~al.}{2013}]{zabalza2013}
{Zabalza} V.,  {Bosch-Ramon} V.,  {Aharonian} F.,   {Khangulyan} D.,  2013,
  \mn@doi [\aap] {10.1051/0004-6361/201220589}, \href
  {http://adsabs.harvard.edu/abs/2013A%26A...551A..17Z} {551, A17}

\makeatother
\end{thebibliography}

\appendix
\section{\boldmath{$T_{\rm sys}$} corrections in the GMRT data}\label{app:tsys}

The power measured by a radio antenna contains the radio emission from a wide number of sources. It is common to separate this power in two components: $P \propto T_{\rm a} + T_{\rm sys}$, where $T_{\rm a}$ is the {\em antenna temperature}, which includes the contribution from the target sources, and \tsys is the {\em system temperature}. \tsys includes the contributions from the Galactic diffuse emission ($T_{\rm sky}$ or sky brightness temperature), the receiver noise, feed losses, spillover and atmospheric emission, all of them considered as the effective receiver temperature, $T_{\rm erc}$ \citep{crane1989,sirothia2009}:
\begin{equation}
	T_{\rm sys}=T_{\rm sky} + T_{\rm erc} \label{eq:tsys}
\end{equation}
While at high frequencies the contribution of the Galactic diffuse emission, $T_{\rm sky}$, is negligible compared to $T_{\rm erc}$, it becomes dominant at low frequencies, thus producing very high system temperatures. This is particularly relevant when observing close to the Galactic Plane. 

To recover the emission from the target sources, \tsys must be determined and removed from the measured power $P$. For single-dish antennas and most interferometers, \tsys can be accurately determined through injected calibration signals and is automatically subtracted for each antenna. 
In the case of GMRT, the observations are, in general, conducted with the Automatic Gain Controller (AGC) enabled to avoid possible limitations such as saturation produced by RFI or excess power. However, this rescales the gain of the system in presence of strong emission (such as the one of the Galactic Plane).
In these kind of observations, the determination of \tsys is not accurate enough and it requires to be accounted for during the calibration of the radio data. The amplitudes of the complex visibilities are proportional to the ratio of the antenna temperature to the system temperature ($T_{\rm a}$/\tsys). In a typical radio observation, the amplitude calibrator allows us to recover the real flux densities of the sky and the target source. In the case of GMRT we will end up with the following visibilities for the calibrator (C) and the target source (S):
\begin{equation}
	V_{\rm C} \propto \frac{T_{\rm a} ({\rm C})}{T_{\rm sys} ({\rm C})},      \qquad V_{\rm S} \propto \frac{T_{\rm a} ({\rm S})}{T_{\rm sys} ({\rm S})}
\end{equation}
During the calibration process, the amplitudes of the complex visibilities are correctly scaled (or calibrated) to flux densities by applying a scaling factor $G\propto T_{\rm sys}({\rm C})$. The scaling factor is determined with the amplitude calibrator, which presents a known constant flux density along the time. Therefore, in this process the contribution of $T_{\rm sys}({\rm C})$ is evaluated and properly removed.
Schematically, after the calibration process we recover the calibrated visibilities, $\tilde{V}$, of the amplitude calibrator:
\begin{equation}
	\tilde{V}_{\rm C} = G\, V_{\rm C}
\end{equation}
However, for the target source we apply the same scaling factor:
\begin{equation}
	\tilde{V}_{\rm S} = G\, V_{\rm S}
\end{equation}
which will only provide properly calibrated visibilities for the target source, $\tilde{V}_{\rm S}$, if $T_{\rm sys}({\rm S}) = T_{\rm sys}({\rm C})$.
For target sources at high Galactic latitudes and/or at high frequencies the Galactic diffuse emission is negligible, and thus \tsys is similar for C and S\footnote{Provided that other effects dependent on the antenna position, such as elevation, are considered during the calibration process.}.
For target sources close to the Galactic Plane and/or at low frequencies
$T_{\rm sys}({\rm S}) \gg T_{\rm sys}({\rm C})$ (because the amplitude calibrator is never located close to the Galactic Plane). In this case, we will need to quantify the ratio $T_{\rm sys}({\rm S}) / T_{\rm sys}({\rm C})$, or \tsys correction, to obtain the correct flux density values in the target source field.
Since LS~5039 lies very close to the Galactic Plane, an accurate correction of \tsys is mandatory to properly estimate the flux densities of the source at each frequency.

\subsection{The Haslam approximation}

A common method to implement this correction is based on estimating the sky temperature values, $T_{\rm sky}$, from the measurements made by the all-sky 408~MHz survey, which was conducted with several single-dish radio telescopes \citep{haslam1982}.

Assuming that the Galactic diffuse emission follows a power-law spectrum, we can determine the sky temperature at a frequency $\nu$ for any position in the sky ($\alpha,\delta$) from the emission reported at 408~MHz by
\begin{equation}
    T_{\mathrm{sky}}^{\,(\nu)} (\alpha,\delta) = T_{\mathrm{sky}}^{\mathrm{(408)}} (\alpha,\delta)\,\left(\frac{\nu}{\mathrm{408~MHz}}\right)^{\gamma}\label{eq:tsky}
\end{equation}
The spectral index for the Galactic diffuse emission is usually assumed to be $\gamma = -2.55$ \citep{roger1999}.
The sky brightness at the positions of the amplitude calibrator and the target source, convolved with the synthesized beam of our interferometer, allows us to estimate the ratio of the sky temperatures between both positions. From eq.~(\ref{eq:tsys}) we have seen that \tsys can be divided in two terms, with $T_{\rm sky}$ being dominant at low frequencies and/or close to the Galactic Plane. The $T_{\rm erc}$ values can be obtained from the tabulated data available in the specification documents of GMRT and the $T_{\rm sky}$ values from eq.~(\ref{eq:tsky}). With these data we determine the $T_{\rm sys}({\rm S}) / T_{\rm sys}({\rm C})$ ratio, which is the \tsys correction.

However, this method has some problems. First, it assumes a constant spectral index for the Galactic diffuse emission across all the sky. Second, it assumes that the response of our instrument to the Galactic diffuse emission is the same as those of the radio telescopes used for the Haslam survey. Third, it
does not take into account the \tsys dependence on the elevation of the Galactic diffuse emission.

\subsection{Direct measurement of the self-power for each antenna}

A more accurate method to implement the \tsys correction should only involve measurements conducted with the same radio telescope.
In the case of an interferometer, one can obtain the power measured at any given frequency and elevation for each antenna in the array by self-correlating the corresponding data from that antenna.
Although the sky brightness is the same for all the antennas, the power that comes from $T_{\rm erc}$ is different for each antenna, given its internal origin. Thus, considering the full array as a collection of isolated single-dish antennas, the problem is reduced to determine the power of each antenna separately, the so-called self-power, and then average all the self-powers to derive the power of the interferometer.

The self-correlated data of each antenna is proportional to the system temperature. Therefore, the ratio between the self-powers for each antenna for the target source and the amplitude calibrator is equal to the ratio between the system temperatures at these two positions. This ratio is thus a more accurate measurement of $T_{\rm sys}({\rm S}) / T_{\rm sys}({\rm C})$ than the Haslam method to estimate the contribution from the Galactic diffuse emission.

As GMRT can not produce self-correlated data at the same time than standard correlated data, additional observations have to be conducted. The self-power for each antenna is measured at the position of the flux calibrator and the position of the source with the same configuration than in the standard correlated observations. 
To avoid possible dependencies with the elevation of the sources, both types of observations should be conducted when the sources are at similar elevations.

The self-correlated data shows stronger RFI than the standard correlated data, and thus a large number of channels must be removed through a dedicated flagging process.
The obtained amplitudes in different channels show a large dispersion. For that reason, it is recommended to take first the average of the ratios for each antenna and each channel, to minimise the dispersion, and then average all the ratios. The use of the median, instead of the mean, is also recommended and it has been used along our data reduction process. The derived value of the ratio is the \tsys correction that must be applied to the image of the target source field to obtain the correct flux density values.

The main source of error in this method is the large dispersion of all the self-powers measured. This can be reduced by accumulating more data obtained in different epochs, because the Galactic diffuse emission is persistent.

\citet{sirothia2009} determined the \tsys corrections for GMRT with direct measurements of the full sky at 240~MHz, and compared the obtained results with the Haslam approximation at that frequency. Although the two methods report similar values for the whole map, these authors found a high scatter in the comparison of the corrections obtained using both methods, with an rms in these differences of $\sim 56\ \%$. Dependencies on the elevation and diurnal/nocturnal time have also been found.

In the case of our target source, we have found the \tsys corrections ($C$) reported in Table~\ref{tab:tsys-corrections}. We observe similar corrections at 610~MHz, but different ones at 235~MHz (of about 60~\%), although the $C_{\rm H}$ and $C_{\rm SP}$ values are roughly compatible at 3-$\sigma$ level in both cases. At 154~MHz the corrections are very different and clearly incompatible. We have seen that the Haslam approximation overestimates the \tsys correction for the field of LS~5039 at all frequencies. At 235~MHz we have obtained an overestimation by an amount in agreement with the rms observed by \citet{sirothia2009} at 240~MHz. As the direct measurements of the self-powers provide more reliable values of the \tsys corrections, and allow us to compute the corresponding uncertainties, we have used $C_{\rm SP}$ to correct our GMRT data.

\begin{table}
\caption{\tsys corrections ($C$) determined in this work for the position of LS~5039 using the amplitude calibrators 3C~286 and 3C~48 with the Haslam approximation ($C_{\rm H}$) and with the direct measurement of the self-powers ($C_{\rm SP}$). Uncertainties are reported at 1-$\sigma$ level. In this work we have used $C_{\rm SP}$ to correct our data.}
\label{tab:tsys-corrections}
\centering
\begin{tabular}{c@{\hspace{+40pt}}c@{\hspace{+40pt}}c}
	\hline
	$\nu/\ \mathrm{GHz}$ & $C_{\rm H}$ & $C_{\rm SP}$\\
	\hline
	154 & $\sim4.0$ & $1.9\pm0.2$\\
	235 & $\sim3.0$ & $1.9\pm0.3$\\
	610 & $\sim1.6$ & $1.39\pm0.06$\\
	\hline
\end{tabular}
\end{table}

\label{lastpage}

\end{document}